\documentclass[pra,twocolumn,amsmath,amssymb,groupedaddress,superscriptaddress]{revtex4}
\usepackage{graphicx}
\newcommand{\ph}{\textmd{ph}}

\def\be{\begin{equation}}
  \def\ee{\end{equation}}

\begin{document}

\title{Quantum phases of interacting phonons in ion traps}

\author{X.-L. \surname{Deng}}
\email{Xiaolong.Deng@mpq.mpg.de}
\author{D. \surname{Porras}}
\email{Diego.Porras@mpq.mpg.de}
\author{J.~I. \surname{Cirac}}
\email{Ignacio.Cirac@mpq.mpg.de}
\affiliation{Max-Planck-Institut f\"ur Quantenoptik, Hans-Kopfermann-Str. 1, Garching, D-85748, Germany.}
\begin{abstract}
The vibrations of a chain of trapped ions can be considered, under
suitable experimental conditions, as an ensemble of interacting phonons,
whose quantum dynamics is governed by a Bose--Hubbard Hamiltonian.
In this work we study the quantum phases which appear in this system,
and show that thermodynamical properties, such as critical parameters
and critical exponents, can be measured in experiments with a limited number of ions. 
Besides that, interacting phonons in trapped ions offer us the
possibility to access regimes which are difficult to study with
ultracold bosons in optical lattices, like models with attractive
or site--dependent phonon-phonon interactions.
\end{abstract}

\date{\today}

\maketitle
\section{Introduction}
%
The interplay between atomic and many--body physics has
proved to be an exciting research field in the last
years. 
Experiments in atomic physics offer us the possibility to find
experimental realizations of theoretical
models which were first proposed in the context of solid state
physics. 
Many of these models are the key to understanding
a variety of phenomena in real materials such as quantum magnetism or
superconductivity. 
For example cold bosonic atoms in optical lattices  
are a realization of the Bose--Hubbard Model (BHM) in a clean
experimental setup, where one can tune the value of interactions and observe
quantum phase transitions in a controlled way \cite{mott}. 
The main handicap of experiments with optical lattices is the fact
that atoms are separated by optical wavelengths, and thus single
particle addressing with optical means is severely limited by
diffraction effects.

Trapped ions are also an experimental system with potential
applications to the quantum simulation of many--body problems
\cite{PorrasCirac.spin,PorrasCirac.phonon,DengPorrasCirac.spin,Milburn,PorrasCirac.2D}. 
It has
the advantage that internal electronic or vibrational quantum states
can be measured at the single particle level
\cite{Leibfried.review,Meekhof}, since the distance
between ions is large enough to address them individually by optical
means. In particular, we have recently shown that the vibrational
modes of a chain of trapped ions under suitable experimental
conditions follow the quantum dynamics of a BHM \cite{PorrasCirac.phonon}. 
The interaction between
phonons is induced by the anharmonicities of an optical potential,
which can be created by an off--resonant standing--wave.

In this work we present a theoretical study of
interacting phonons in trapped ions, and show the following results:
(i) The quantum phase transition between a superfluid and a Mott
insulator phase can be induced and observed in this system. 
(ii) Even though finite size--effects are important due to the finite
length of the ion chain, properties corresponding to the
thermodynamical limit can be accessed in experiments with a limited
number of ions. These include critical exponents of
correlation functions, and critical values of parameters in the
Hamiltonian. 
(iii) The ability to control phonon--phonon interactions allows us to
realize models of interacting bosons which are difficult to reproduce
in other experimental setups, like for example, BHM's with negative
interactions, as well as models with site dependent interactions.

The structure of the paper is the following. In section 
\ref{section.Bose.Hubbard.model} we derive the Bose--Hubbard model for
phonons in a chain of trapped ions, in the presence of the
anharmonicities induced by an optical
dipole potential. The DMRG algorithm that we have used to study
numerically this problem is summarized in section \ref{section.numerical.method}.
In sections \ref{section.repulsive.interactions} and 
\ref{section.attractive.interactions}, we study the quantum phases
which correspond to repulsive and attractive
phonon--phonon interactions, respectively. Section
\ref{section.site.dependent.interactions} is devoted to the case of a
Bose--Hubbard model with site--dependent interactions. Finally in the
last section we summarize our results and conclusions.

\section{Bose-Hubbard model of phonons in ion traps} 
\label{section.Bose.Hubbard.model}
In this section we show that under certain experimental conditions, the dynamics of
the vibrational modes of a chain of ions satisfies the Bose-Hubbard
model of interacting phonons in a lattice \cite{PorrasCirac.phonon}. 
Phonon number conservation is ensured whenever vibrational
energies are much higher than other energy scales in the system. 
In this limit, physical processes which involve creation or destruction 
of a phonon do not conserve particle number and are suppressed in much
the same way as processes which do not conserve the number of electrons or atoms 
in low--energy physics.

\subsection{Hamiltonian in the harmonic and phonon conserving approximation}

Let us start by writing the Hamiltonian that describes a chain of ions in a linear trap:
\begin{equation}
H_0 = \sum_{i = 1}^N \frac{{\vec{P}}_i^2}{2 m} + V_{\textmd{T}} +
\sum_{\substack{i,j =1 \\ (i > j) }}^N 
\frac{e^2}{|\vec{R_i} - \vec{R_j}|} .
\end{equation}
$N$ is the number of ions, and $m$ is their mass.
$\vec{P}_j$ and $\vec{R}_j$ are the momenta and the absolute positions
of the ions, respectively. $V_{\textmd{T}}$ is the trapping potential,
which determines the ions' equilibrium positions.  
In this work we deal with two different situations. 
On one hand we consider the case of ions in a linear Paul trap, 
where they are confined by an overall trapping potential:
\begin{equation}
V_{\textmd{T}} = \frac{1}{2} m \sum_{i = 1}^N   \sum_{\alpha = x,y,z}
\omega_\alpha^2 R^2_{i,\alpha} ,
\label{Paul.trapping.potential}
\end{equation}
where $\omega_\alpha$ are the trapping
frequencies in each spatial direction.
On the other hand, we consider ions confined by an array of separate microtraps:
\begin{equation}
V_{\textmd{T}} = 
\frac{1}{2} m \sum_{i = 1}^N \sum_{\alpha = x,y,z}
\omega^2_{\alpha} \left(R_{i,\alpha} - \bar{R}_{i,\alpha} \right)^2,
\label{array.trapping.potential}
\end{equation}
where $\bar{R}_{j,\alpha}$ are the centers of each microtrap. Note
that in (\ref{array.trapping.potential}) we assume that the
confinement is strong enough, such that each ion
feels only a single microtrap.

The equilibrium positions of the ions are given by the minima of the
trapping potential plus the Coulomb repulsion. From now on, we choose
the condition $\omega_z \gg \omega_x, \omega_y$, such that the ion
chain is along the ${\bf z}$ axis, with equilibrium positions given by
$z_i^0$. In the case of a linear Paul trap,
described by Eq. (\ref{Paul.trapping.potential}), the equilibrium
positions of the ions have to be calculated numerically. The distance
between ions is smaller at the center of the chain. In the case of
independent microtraps (Eq. (\ref{array.trapping.potential})), 
one can approximate the equilibrium positions
by assuming that they correspond to the center of each microtrap:
$z_i^0 = \bar{R}_{i,z}$. This fact has strong implications for the phonon
quantum dynamics, as we will see below.

In the harmonic approximation, $H_0$ is expanded up to second order in
the displacements of the ions around
the equilibrium positions, and we get a set of independent vibrational
modes corresponding to each spatial direction. 
The phonon number is a conserved quantity if the vibrational energies
are the largest energy scale in the system. 
This condition can be met
either in the case of ions in individual microtraps, or in the case of
the radial vibrations of ions in a linear trap, because the
corresponding trapping frequencies can be increased without destroying
the stability of the ion chain.
For concreteness we
restrict from now on to the case of vibrations in one of the radial
directions, say ${\bf x}$, but keep in mind that our results can be applied also to
the axial vibrational modes if ions are in individual microtraps. 

The Hamiltonian that governs the dynamics of the radial coordinates in the
harmonic approximation reads:
\begin{eqnarray}
H_{\textmd{x0}} = \sum_{i=1}^N \frac{P^2_{i,x}}{2 m} &+& 
\frac{1}{2}  m \omega_x^2 \sum_{i=1}^N x_i^2 \\ \nonumber 
&-& \frac{1}{2} \sum^N_{\substack{i,j = 1 \\ (i > j)}} 
    \frac{e^2}{|z_i^0 - z_j^0|^3} \left(x_i - x_j \right)^2 ,
\label{harmonic.x.1}
\end{eqnarray}
where $x_i$ are the displacements of the ions around the equilibrium
positions, that is, simply $x_i = R_{i, x}$, and $P_{i,x}$ the
corresponding momenta.
The second quantized form of this Hamiltonian is (we consider units
such that $\hbar = 1$):
\begin{equation}
H_{\textmd{x0}} 
= \sum_{i=1}^N \omega_{x,i} a^\dagger_i a_i 
+ 
\sum^N_{\substack{i,j = 1 \\ (i > j)}} t_{i,j} 
\left( a^\dagger_i + a_i \right) \left( a^\dagger_j +  a_j \right).
\label{harmonic.x.2}
\end{equation}
$a_i^\dagger$ ($a_i$) are creation (anihilation) operators for
phonons in the radial direction.
Harmonic corrections induced by the Coulomb interaction determine the
effective trapping frequency, $\omega_{x,i}$ which depends on the ions' positions, as
well as the tunneling amplitudes $t_{i,j}$:
\begin{eqnarray}
\omega_{x,i} &=& \omega_x -  
\frac{1}{2} \sum_{\substack{j = 1 \\ (j \neq i)}}^N 
\frac{e^2/(m \omega_x^2)}{|z_i^0 \! - \! z_j^0|^3} \  \hbar \omega_x, 
\label{define.omegax}
 \\
t_{i,j} &=& \frac{1}{2} \frac{e^2/(m \omega_x^2)}{|z_i^0 \! - \!
  z_{j}^0|^3} 
\ \hbar \omega_x .
\label{define.tx}
\end{eqnarray}
Eq. (\ref{define.omegax}) yields an important result on the properties
of phonons in trapped ions: the corrections to the local trapping
energy, $\omega_{x,i}$ may depend on the position of the ions, 
in case the distance between ions changes along the chain. In 
\cite{PorrasCirac.phonon} we have shown that in a linear ion trap,
ions arrange themselves in a Coulomb chain, such that $\omega_{x,i}$
is an effective harmonic confining potential for the phonons. On the
contrary, in the case of an array of ion microtraps, the distances between ions
can be considered to be approximately constant, and thus this
confining effect does not take place.

Before going any further, let us study under which conditions phonon
nonconserving terms of the form ($a_i a_j$, $a_i^\dagger a_j^\dagger$),
can be neglected in Eq. (\ref{harmonic.x.2}). 
We define the parameter: 
\begin{equation}
\beta_x = e^2/(m \omega_x^2 d^3_0),
\end{equation} 
where $d_0$ is the distance between ions. Since we will be interested
in the limit $\beta_x \ll 1$, we choose $d_0$ to be the minimum
distance between ions in the case of a linear Paul trap. Phonon
tunneling terms $t_{i,j}$ are of the order of $t$, defined by:
\begin{equation}
t = \beta_x \omega_x / 2 .
\label{define.t}
\end{equation}
Since phonon nonconserving terms rotate fast in (\ref{harmonic.x.2}),
we can neglect them in a rotating wave approximation if:
\begin{equation}
t/\omega_x =  \beta_x/2 \ll 1,
\end{equation}
such that Eq. (\ref{harmonic.x.2}) takes the form of tight binding
Hamiltonian with hopping terms 
$t_{i,j} ( a_i^\dagger a_j + h.c )$.

In our numerical calculations, we will parametrize the tunneling of
phonons between sites by the parameter $t$, which corresponds, due to
the definition of $\beta_x$, to the highest value of the tunneling
along a chain in the case of a linear Paul trap, and to the tunneling
between nearest--neighbors in the case of an array of microtraps.

\subsection{Phonon--phonon interactions}

Anharmonic terms in the vibrational Hamiltonian are interpreted as phonon--phonon
interactions, and can be induced by placing the ions at the minimum or
maximum of the optical dipole potential created by an off--resonant
standing wave along ${\bf x}$:
\begin{equation}
H_{\textmd{sw}} = F \sum_{i=1}^N \cos^2 (k x_i + \frac{\pi}{2} \delta) .
\label{standing.wave.1}
\end{equation}
$F$ is the amplitude of the dipole potential, and $\delta$
determines the position of the ions relative to the standing wave. 
We define the Lamb--Dicke parameter $\eta = k x_0$, where $k$ is the
wave--vector of the standing wave lasers, and $x_0$ is the
ground--state size of the radial trapping potential. The only relevant
cases for us are $\delta = 0$ (maximum of the optical potential), and
$\delta = 1$ (minimum). 
Under the condition
$\eta \ll 1$, we can write $H_{\textmd{sw}}$ as a series around 
$x_i = 0$. 
The term that is quadratic in $x_i$ in
Eq. (\ref{standing.wave.1}) can be included in the harmonic
vibrational Hamiltonian just by redefining the global radial trapping frequency:
\begin{equation}
\omega^2_x \rightarrow \omega_x 
\left( \omega_x - (-)^\delta 4 \eta^2 F \right) .
\label{frequency.shift}
\end{equation}
In the case $\delta = 0$, condition $\eta^2 F \ll \omega_x$ has to be
fulfilled, such that the radial trapping frequency is not strongly
suppressed by the standing wave, and the system remains in the phonon
number conserving regime. Under this condition the only relevant
term is thus the quartic one:
\begin{equation}
H^\textmd{(4)}_\textmd{sw} = 
(-1)^\delta \frac{F \eta^4}{3} 
\sum_{j=1}^N \left(a_j + a_j^\dagger \right)^4 .
\label{quartic}
\end{equation}
We can neglect nonconserving phonon terms again under the condition
$F \eta^4 \ll \omega_x$. In this way we get, finally, the promised
BHM for phonons:
\begin{eqnarray}
H^{\textmd{BHM}}_{\textmd{x}}
&=&
\sum^{N}_{\substack{i,j = 1 \\ i > j}} t_{i,j}(a^{\dagger}_i a_j+h.c.) \nonumber \\
&+&
\sum^{N}_{i=1}(\omega_x+\omega_{x,i}) a^{\dagger}_i a_i +
U \sum^{N}_{i=1} a^{\dagger2}_i a^2_i\label{Hamiltonian.phonon} .
\label{BHM}
\end{eqnarray}
The on--site interaction is given by:
\begin{equation}
U = 2 (-1)^\delta F \eta^4 . 
\end{equation}
Thus, it is
repulsive or attractive depending on whether the ions are placed at a minimum or maximum
of the standing-wave, respectively.

\section{Numerical Method}
\label{section.numerical.method}
The Bose--Hubbard Model with tunneling between nearest--neighbors has
been thoroughly studied in the past \cite{Fisher,Sachdev}. 
It has recently received considerable attention 
because it describes experiments with ultracold atoms in optical
lattices. 
In general, we expect the same phenomenology to appear in our
problem, such as, for example, a superfluid--Mott insulator quantum phase
transition. 
However, the situation of phonons in ion traps presents a few
peculiarities that deserve a careful analysis: 
the effects of long--range tunneling in (\ref{BHM}), finite size
effects, as well as the possibility of having
attractive interactions.

To handle this many--body problem numerically we use the Density Matrix
Renormalization Group method (DMRG) \cite{White_DMRG}, 
which has proved to be a quasi--exact method in
quantum chains. 
In particular we use the finite-size algorithm for open
boundary conditions. Our problem is defined in the microcanonical
ensemble, that is, we find the minimum energy state
within the Hilbert subspace with a given number of total phonons
$N_{\textmd{ph}}$.
For this reason, we have implemented a DMRG code which uses the total
phonon number as a good quantum number and projects the problem into
the corresponding subspace at each step in the algorithm \cite{Schollwoeck_Review}.

To keep a finite dimensional Hibert space, we truncate the number of
phonons in each ion, and define a maximum value
$n_{max}$, which is usually taken to be of the order of $6 \langle n
\rangle$, with $\langle n \rangle$ the mean phonon number. 
The number of eigenstates of the reduced density matrix that we keep
at each step is always in the range $80 - 100$.

We have checked the accuracy of our method by comparing our numerical
calculations with the exact solution in the case of a system of
non--interacting ($U = 0$) phonons, where the ground state at zero
temperature is a condensate of phonons in the lowest energy
vibrational state. We have also compared our numerical results with
exact diagonalizations of Eq. (\ref{BHM}) with up to $N=5$ ions. In
both cases we found agreement between DMRG and the exact results up to
machine accuracy $\delta E \sim 10^{-14}$.   

The relevant experimental parameters of our phonon-Hubbard model are discussed
in the Ref. \cite{PorrasCirac.phonon}. Typically we could choose the minimum distance
between ions $d_0 = 5$  $\mu m$ and $\beta_x = 2\times 10^{-2}\ll 1$. Then we
have $\omega_z \approx 177$ $\textrm{kHz}$ and $\omega_x \approx 12.5$
$\textrm{MHz}$
for a string of ions with $N=50$. The number of phonons is $N_{ph} = N$ for
the superfluid and Mott-insulator phases and $N_{ph} = N/2$ for the Tonks-gas
phase.  In the end we discuss $N_{ph} = 2N$ for a special case with
site-dependent interactions. All our calculations are for the ground state,
i.e. at zero temperature. 

Following the discussion below Eqs. (\ref{define.omegax},
\ref{define.tx}), one expects to find significant differences between
the cases of phonons in ions trapped in a linear trap (Coulomb chain),
and phonons in an array of ion microtraps. Finite size and
inhomogeneity effects are indeed much more important in the linear
trap case, since harmonic Coulomb corrections induce an effective
harmonic trapping for the phonon field. For this reason, we always
study these two cases separately, in the different quantum phases that
we will explore in what follows.

\section{Bose-Hubbard Model with Repulsive interactions: $U>0$}
\label{section.repulsive.interactions}
We study first the quantum phases of phonons with $U > 0$, 
and both commensurate and incommensurate total phonon number.
In this section we present results for a chain with $N=50$ ions, 
and total phonon number $N_{\textmd{ph}} = N$ in the commensurate
case, or $N_{\textmd{ph}}=N/2$ in the incommensurate case.

The local observables that we consider are the number of phonons
at each site, $ n_j = \langle a^{\dagger}_j a_j\rangle$, 
as well as its fluctuations,  $\delta n_j = 
\sqrt{\langle  n^2_j\rangle - \langle n_j \rangle^2}$. 
Two--point correlation functions have to be defined carefully, to take
into account finite size effects, for example, the variations of the
density of phonons along the chain.
Correlations in the number of phonons are given by:
\begin{eqnarray}
C^{nn}_{i,j} = \langle n_i n_j\rangle - \langle n_i\rangle\langle n_j\rangle.
\end{eqnarray}
A suitable definition of correlations that are non--diagonal in the phonon
number basis is the following one \cite{Kollath_correlations}:
\begin{eqnarray}
C^{aa}_{i,j} = \frac{\langle a^{\dagger}_i a_j \rangle }
{\sqrt{\langle n_i \rangle \langle n_j \rangle}},
\label{nondiagonal.correlations}
\end{eqnarray}
such that correlations are rescaled by local values of the phonon
density. The rescaling is inspired by the decomposition of the
phonon field in density and phase operators, $a_j = \sqrt{n_j} e^{-i
  \phi_j}$, which is the starting point for the Luttinger theory of the weakly
interacting bosonic superfluid \cite{Haldane,Gangardt,Kheruntsyan}.

\begin{figure}
\resizebox{1.68in}{!}{\includegraphics{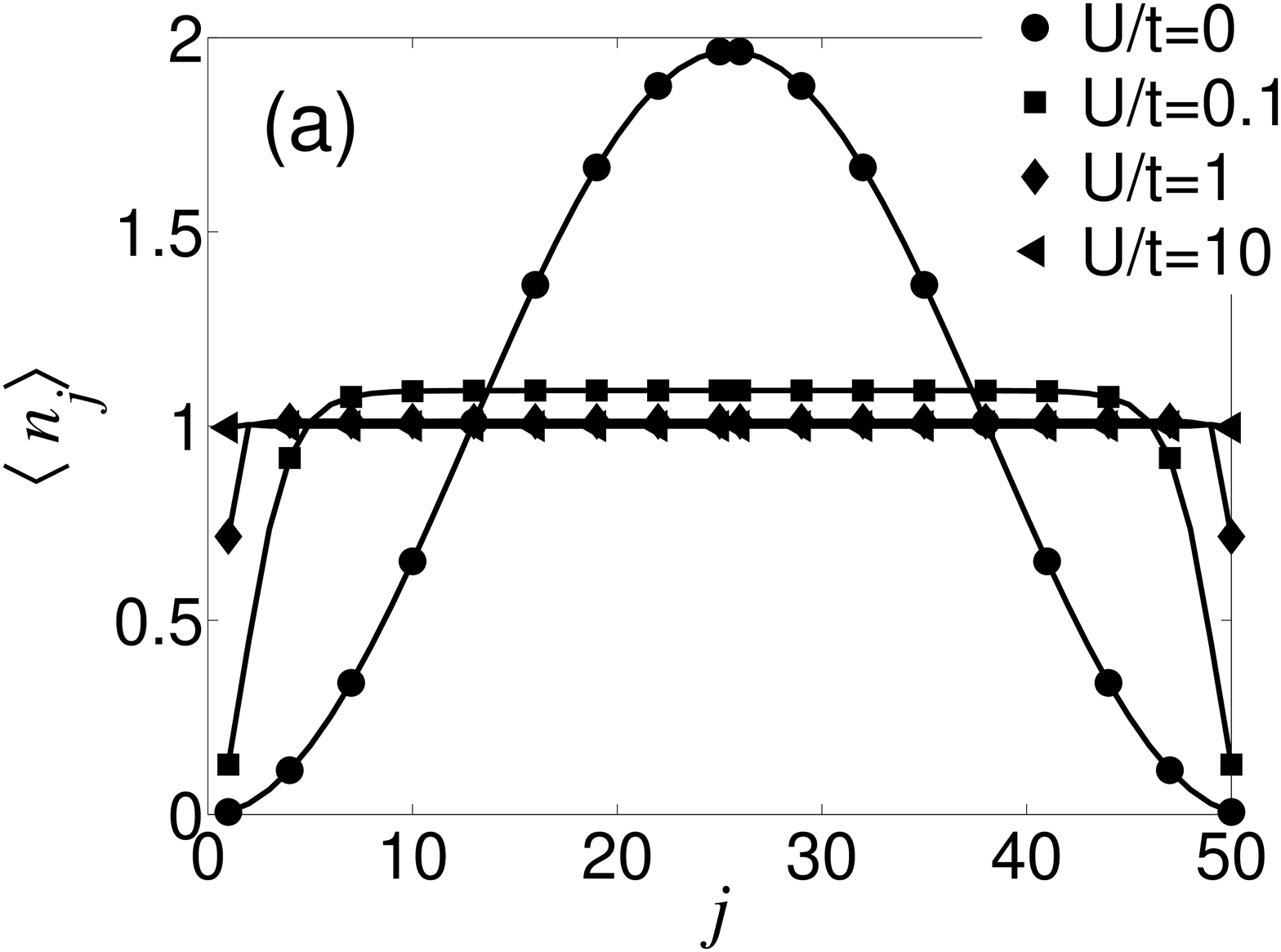}}
\resizebox{1.68in}{!}{\includegraphics{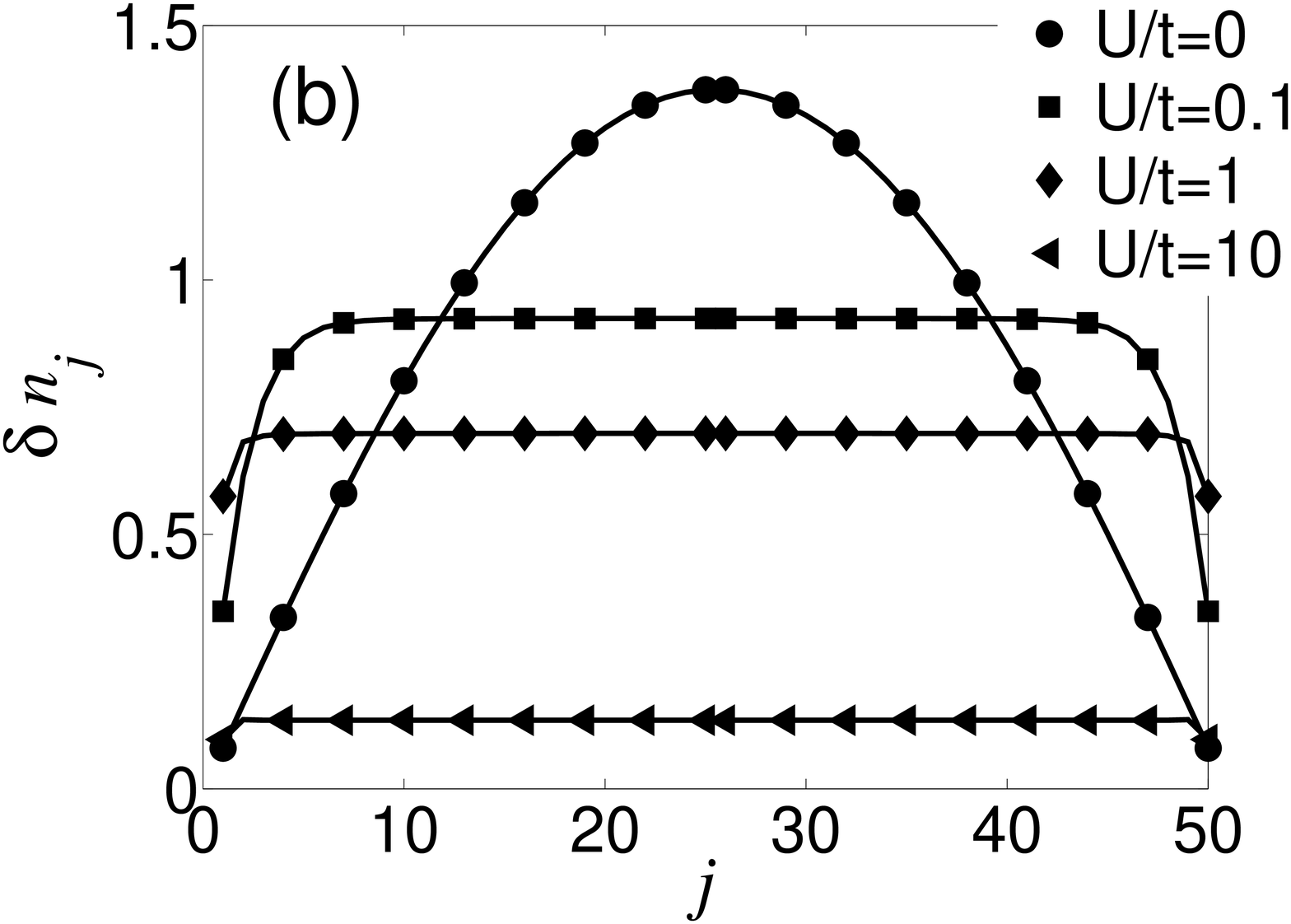}}
\caption{(a) Mean phonon number $\langle n_j \rangle$, and (b) fluctuations
  $\delta n_j$, at each ion in the phonon ground state in an array of
  microtraps. Number of ions $N=50$, and total phonon number $N_{\textmd{ph}}=50$.}
\label{DensityFluctuation_Micro}
\end{figure}
\begin{figure}
\resizebox{1.68in}{!}{\includegraphics{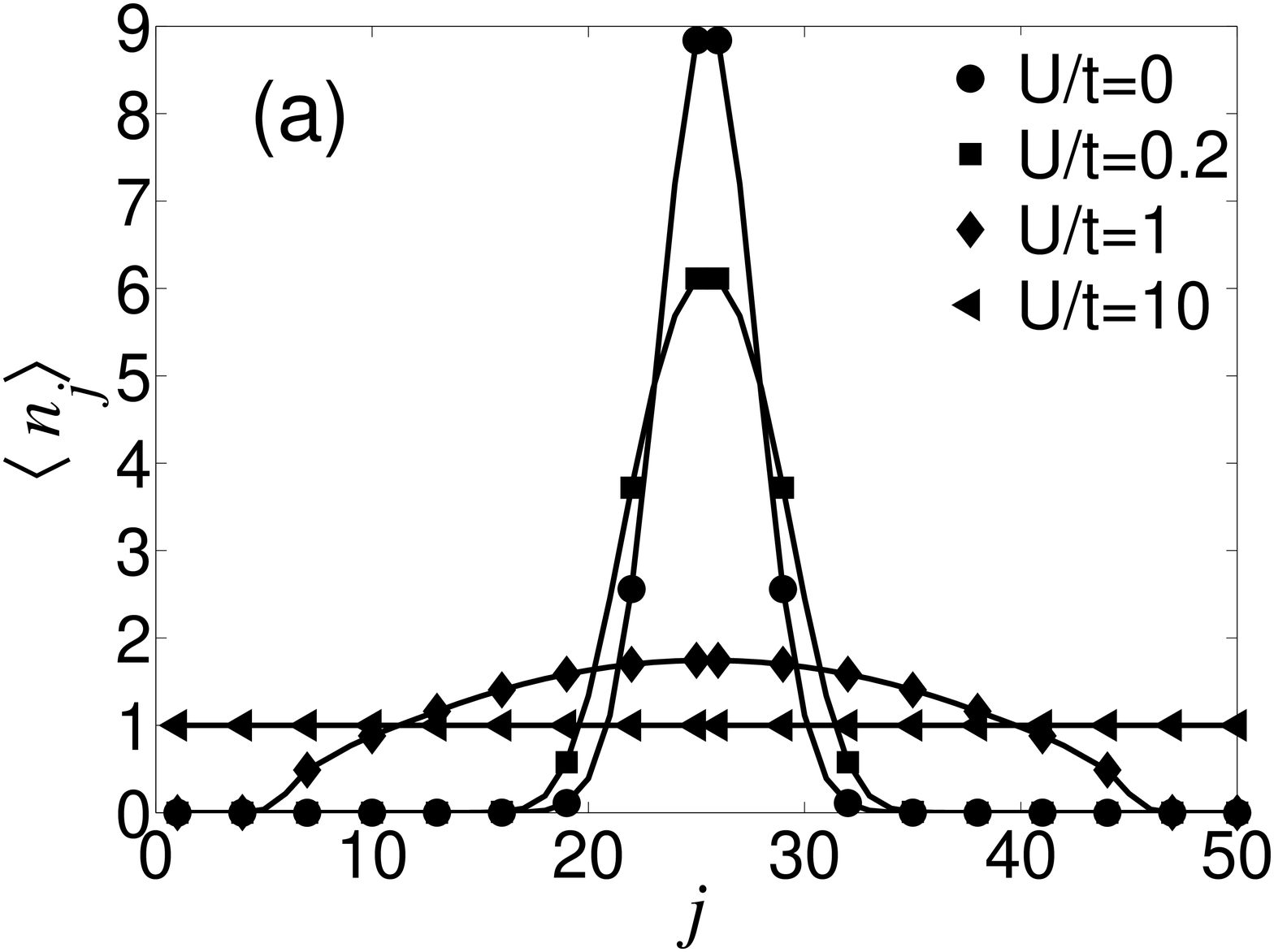}}
\resizebox{1.68in}{!}{\includegraphics{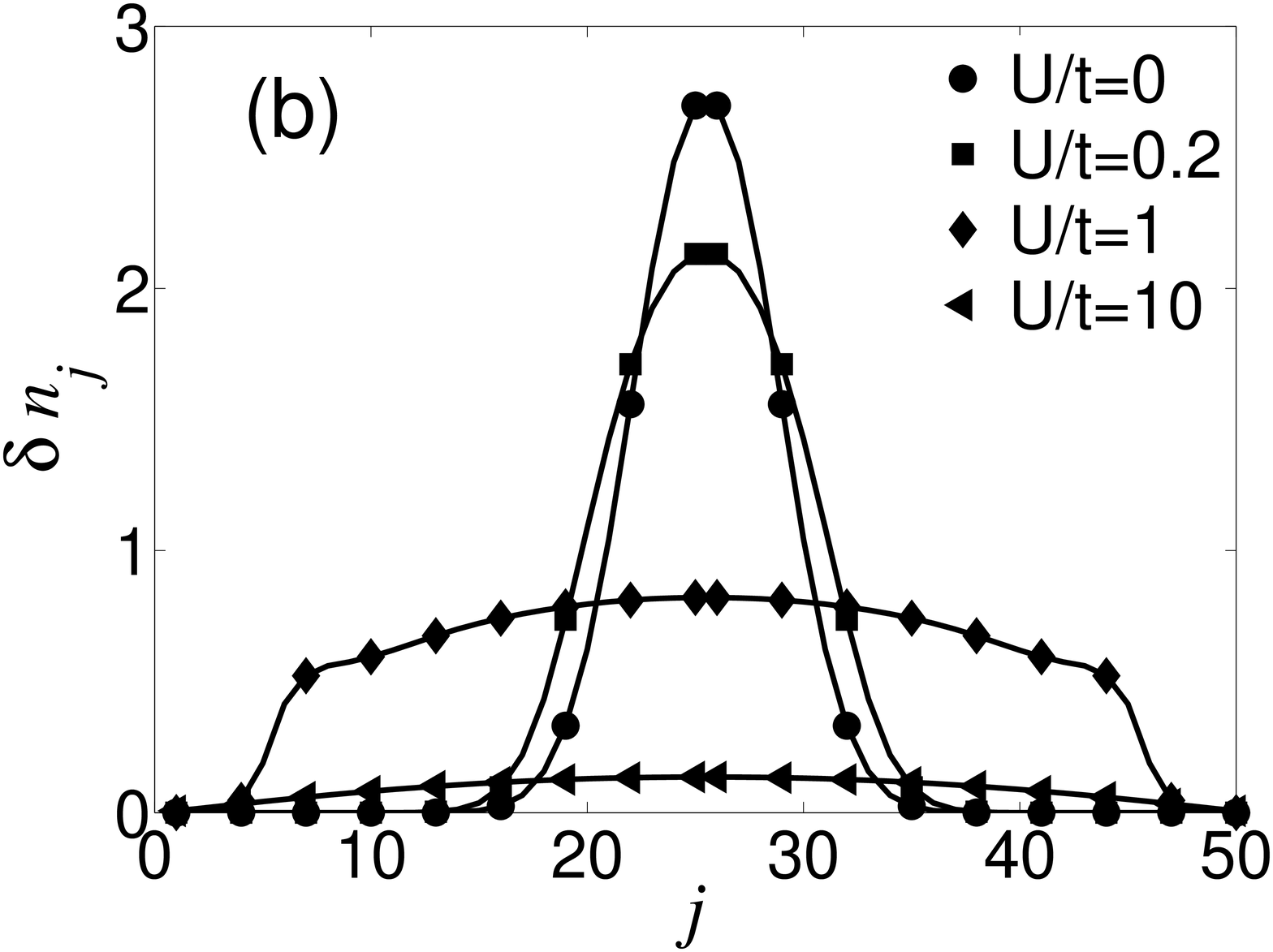}} 
\caption{(a) Mean phonon number and (b) fluctuations, at each 
ion in the phonon ground state in a
linear Paul trap. $N = 50$, $N_{\textmd{ph}} = 50$.}
\label{DensityFluctuation_Paul}
\end{figure}

In Figs. \ref{DensityFluctuation_Micro} and \ref{DensityFluctuation_Paul}, 
we plot the local density and its fluctuations in the cases of an
array of microtraps, and a linear Paul trap, respectively.
These figures show a signature of the different phases which can be observed in the model
defined by the Hamiltonian (\ref{BHM}). 
Figs. \ref{DensityFluctuation_Micro}(a) and
\ref{DensityFluctuation_Paul}(a), in particular, show the variation
of the density of phonons along the chain.
The evolution of the density profile shows the 
transition from the phonon superfluid to the
Mott--insulating phase. When $t \gg U$, the ground state of the system is a
condensate such that all the phonons occupy the lowest energy
vibrational mode. In the case of a linear Paul trap,  phonons are 
confined in the center of the chain, due to the effective trapping
potential induced by the nonconstant ion--ion distance. 
At $U \gg t$, the ground state is a phonon Mott insulator with
approximately one phonon per site, and no phonon number fluctuations.
Note that due to the effective harmonic trapping potential, the Mott
phase in the whole chain is reached for lower values of $U$ in the
case of the array of microtraps (Fig. \ref{DensityFluctuation_Micro})
than in the linear Paul trap case (Fig. \ref{DensityFluctuation_Paul}).

In the following subsections, we study these two quantum phases
separately, paying particular attention to their correlation functions. 

\subsection{Superfluid phase} 
When the tunneling dominates the on--site interactions, the system
is in the superfluid phase \cite{note1}. The non-interacting ground state is given
by a condensate solution in which the $N_{\textmd{ph}}$ phonons are in
the lowest vibrational mode:
\begin{equation}    
| \psi_{SF} \rangle = 
\frac{1}{\sqrt{N_\ph !}}
\left(
\frac{1}{\sqrt{N}} \sum_i {\cal M}^0_{i} a^{\dagger}_i \right)^{N_\ph} | 0 \rangle ,
\label{ground.state.condensate}
\end{equation}
where ${\cal M}^0_i$ is the wave--function of the lowest energy vibrational mode.
Interactions suppress long range order in 1D, even in the weak
interacting limit, $U \ll t$, in which Luttinger liquid theory
allows us to make predictions on the scaling of correlation functions:
\begin{eqnarray}
C^{aa}_{i,j} \propto |i-j|^{-\alpha}, \nonumber \\
C^{nn}_{i,j} \propto |i-j|^{-2} ,
\label{luttinger.scaling}
\end{eqnarray}
where $\alpha$ depends on the parameters of the model:
\begin{eqnarray}
\alpha \propto \sqrt{\frac{U/t}{n_0}} .
\label{luttinger.scaling.alpha}
\end{eqnarray}
In deriving (\ref{luttinger.scaling}) one has to neglect phonon tunneling
beyond nearest--neighbor ions, and assume an homogeneous
system \cite{Haldane}. In the following we will check if Luttinger theory describes
also our numerical results in the case of phonons in a chain of
trapped ions, by fitting our results to the form
(\ref{luttinger.scaling}).

\begin{figure}
\resizebox{1.65in}{!}{\includegraphics{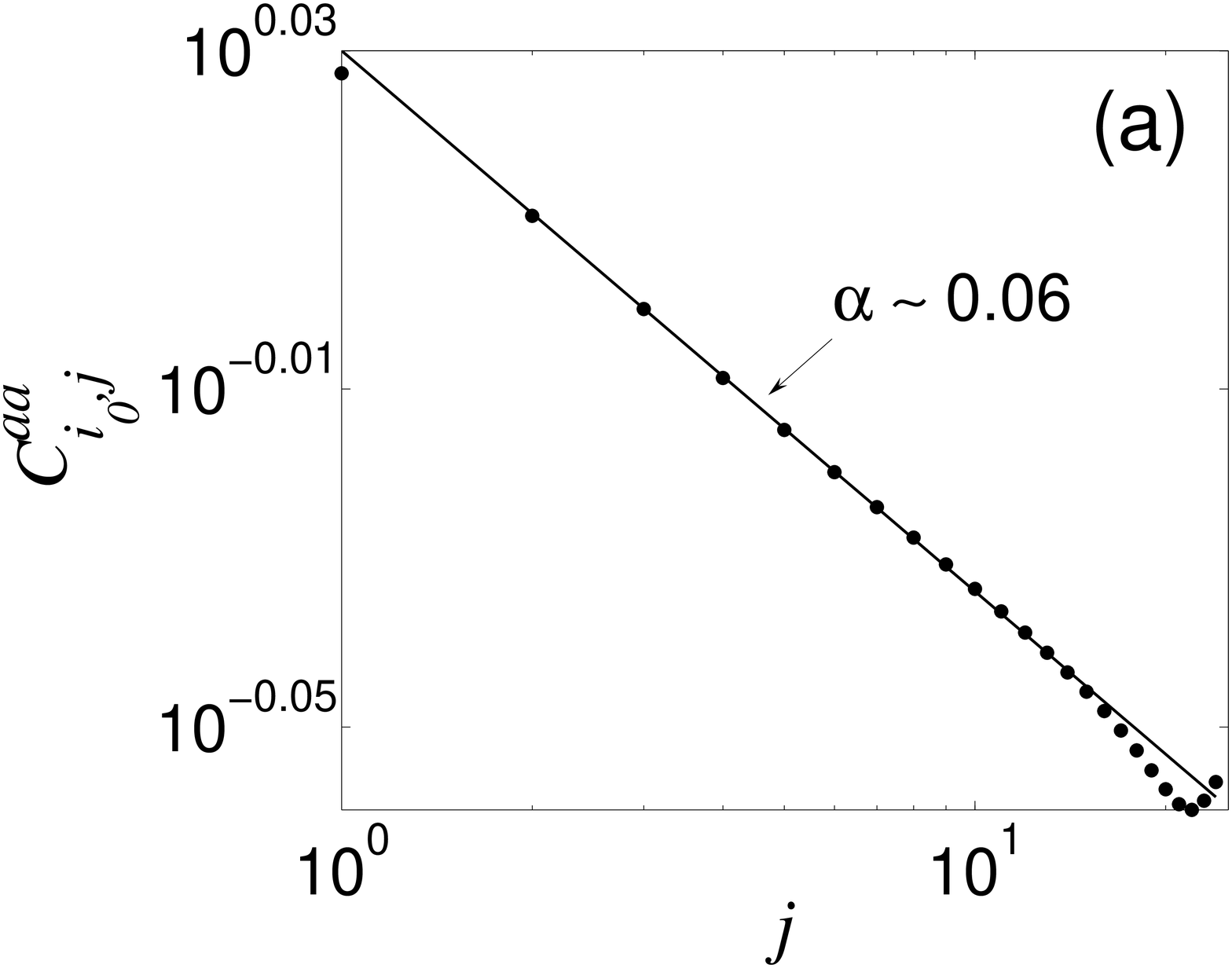}}
\resizebox{1.65in}{!}{\includegraphics{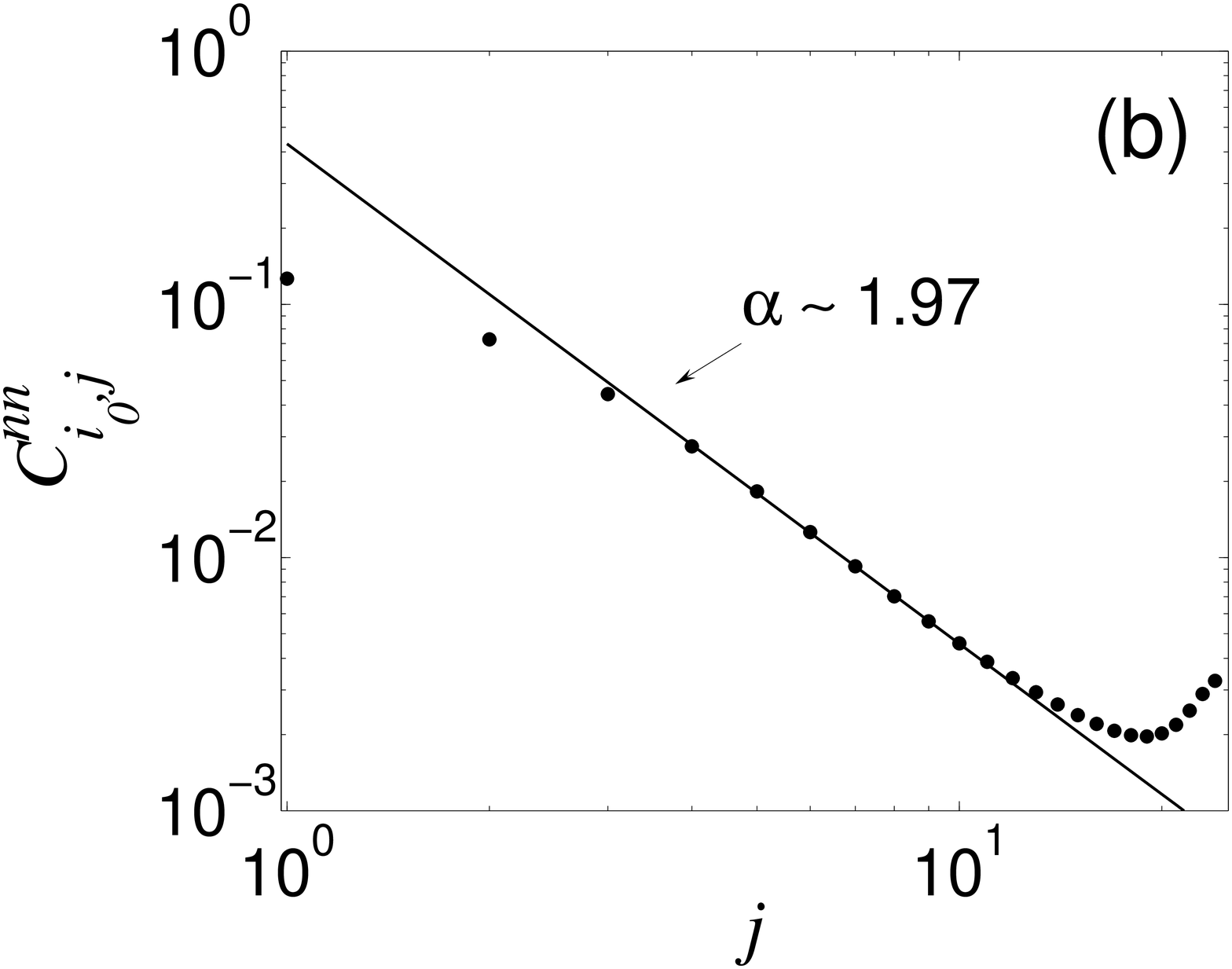}}
\caption{Correlation functions (a) $C^{aa}_{i_0,j}$ and (b) $C^{nn}_{i_0,j}$ as a function of
  coordinate $j$, at the superfluid phase  ($U/t=0.1$) in  an array of
ion  microtraps. We choose $i_0 = 26$ (center of the chain), $N = N_\textmd{ph} = 50$.
The dotted and solid lines are numerical results and fittings in the
region where the functions show algebraic decay.}
\label{MicroCorrSF}
\end{figure}
\begin{figure}
\resizebox{3.0in}{!}{\includegraphics{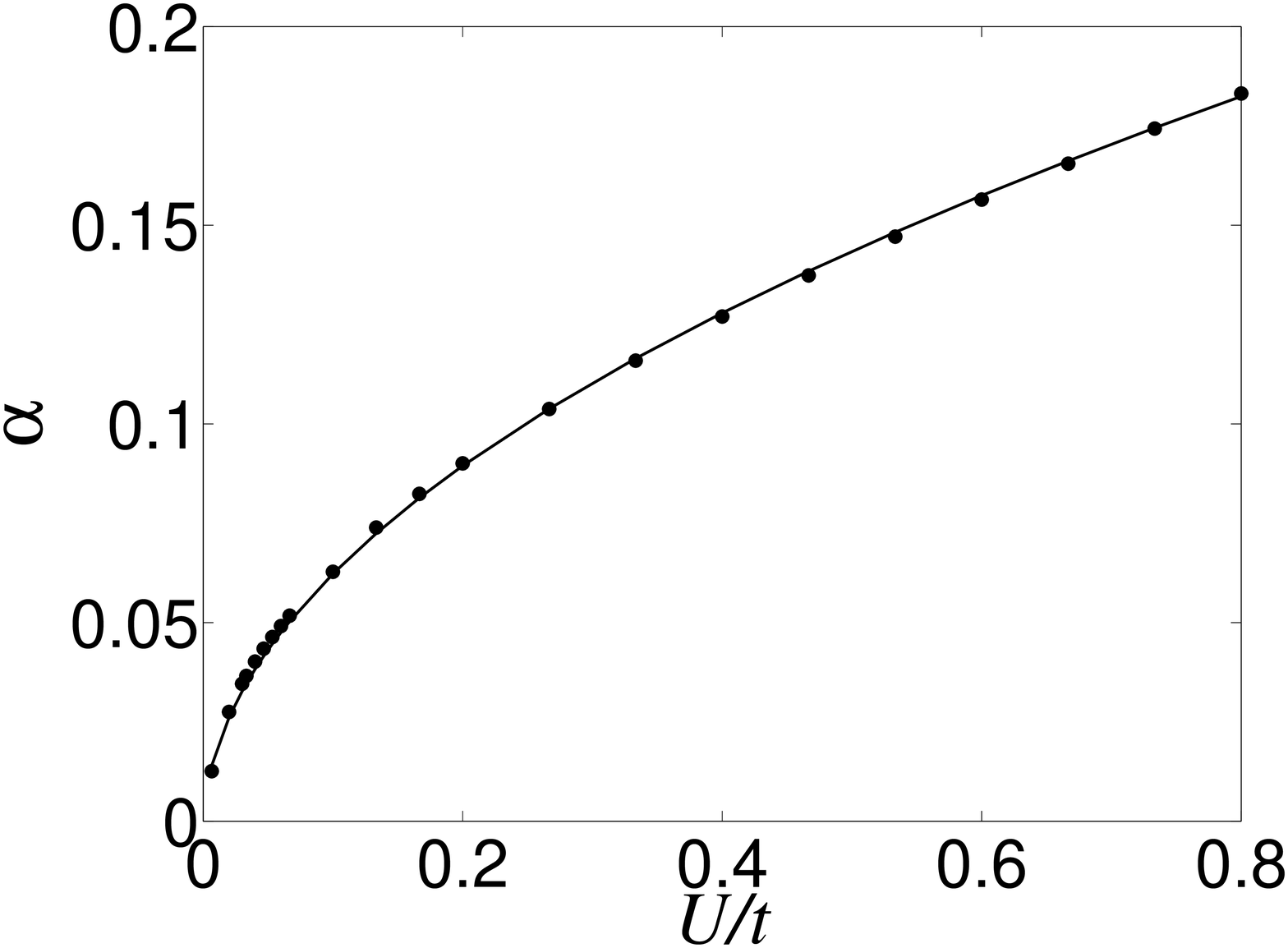}}
\caption{The evolution of the
  exponent $\alpha$ of $C^{aa}_{i_0,j}$ as a function of $U/t$ in the regime of
  superfluid phase, in an array of ion microtraps ($i_0 = 26$, $N =
  N_\textmd{ph} = 50$).
The dotted line is from numerical data, and the solid line the
  fitting result from the Luttinger liquid, that is, $\alpha \approx
  A\sqrt{\frac{U/t}{n_0}}$, where the coefficient $A\approx 1.68$.}
\label{MicroCorrEvol}
\end{figure}

{\it (1) Array of microtraps.}
We start with the case of the superfluid phase in an array of ion
microtraps, see Fig. \ref{MicroCorrSF}. 
Correlation functions $C^{aa}_{i,j}$ and $C^{nn}_{i,j}$ 
decay algebraically in an intermediate range of
ion--ion separations, with exponents which satisfy the predictions of
Luttinger theory. In particular, the evolution of
$\alpha$ in Eq. (\ref{luttinger.scaling})
is well described by the Luttinger liquid scaling law
(\ref{luttinger.scaling.alpha}), 
as shown in Fig. \ref{MicroCorrEvol}. 

\begin{figure}
\resizebox{1.68in}{!}{\includegraphics{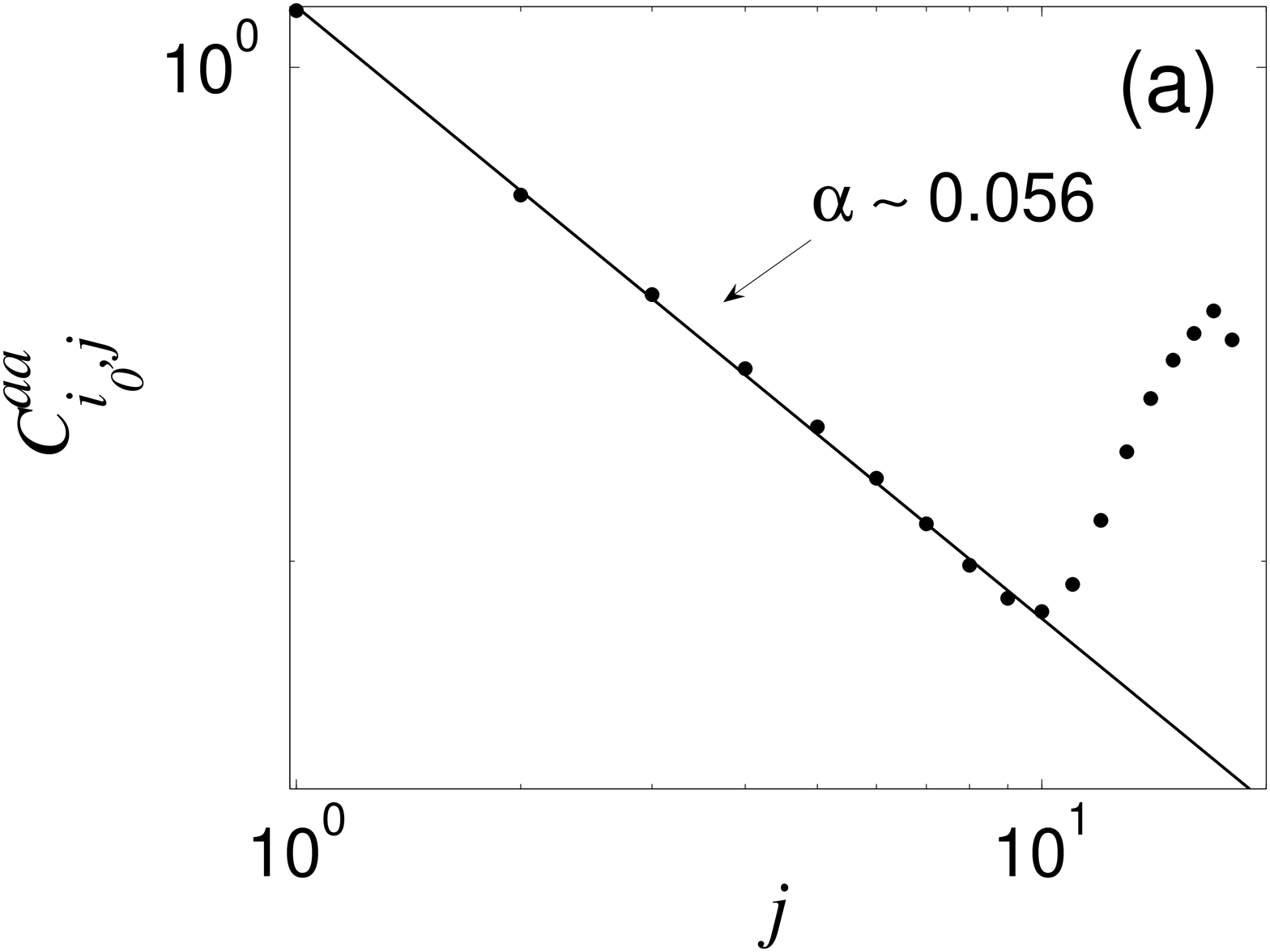}}
\resizebox{1.68in}{!}{\includegraphics{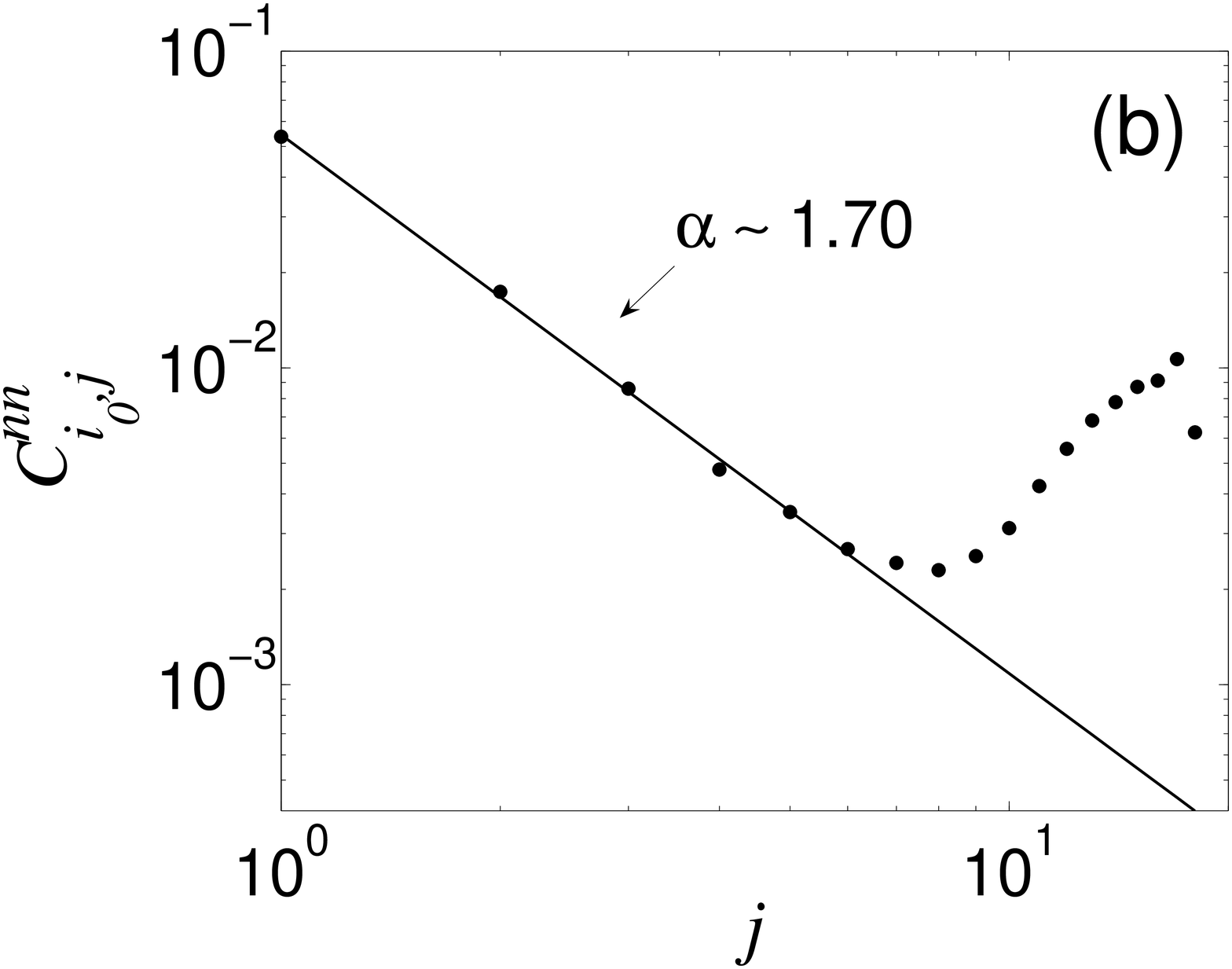}}
\caption{Correlation functions (a) $C^{aa}_{i_0,j}$ and (b) $C^{nn}_{i_0,j}$ at the
  superfluid phase $U/t=0.2$ in a linear Paul trap 
($N = N_\textmd{ph} = 50$). The dotted lines are
numerical data, and the solid lines are fittings in the region where
the correlations decay algebraically.}
\label{PaulCorrSF}
\end{figure}

\begin{figure}
\resizebox{3.0in}{!}{\includegraphics{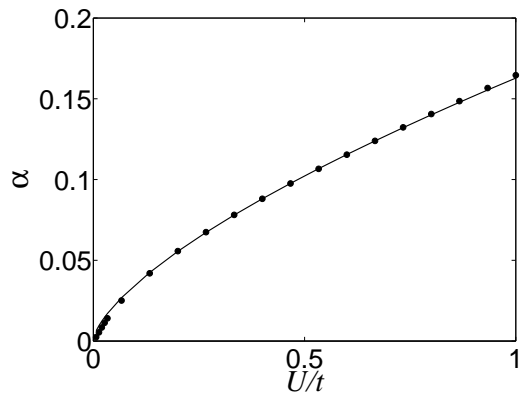}}
\caption{The evolution of the parameters $\alpha$ in $C^{aa}_{i_0,j}$ with
  $U/t$. 
The dotted line is the values from the correlation, and the solid
  line is the fitting data from $\alpha \approx 0.215\sqrt{\frac{U/t}{n_0}}$, corresponding to the
  Luttinger liquid theory.}
\label{PaulCorrEvol}
\end{figure}

{\it (2) Linear ion trap.}
In the case of ions in a linear Paul trap, finite size effects play a
more important role, because of the inhomogeneities of the on--site
phonon energy. Correlation functions
still decay algebraically for short distances in the superfluid
regime, but boundary effects spoil this behavior at large separations between ions. 
In the algebraic regime, exponents are close to those predicted by
Luttinger theory in the homogeneous case, see Figs. \ref{PaulCorrSF} and
\ref{PaulCorrEvol}. 

\begin{figure}
\resizebox{3.0in}{!}{\includegraphics{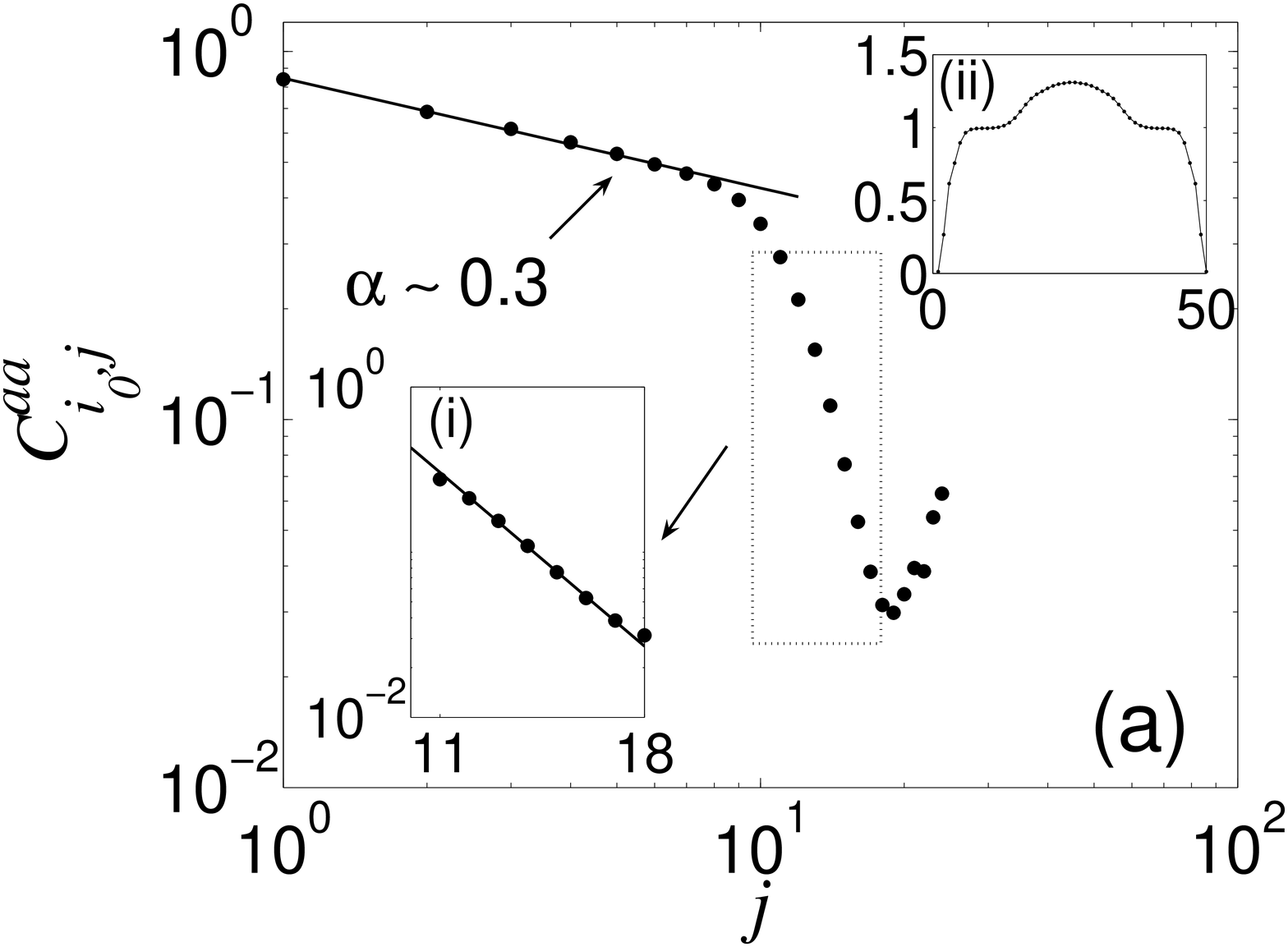}}
\resizebox{3.0in}{!}{\includegraphics{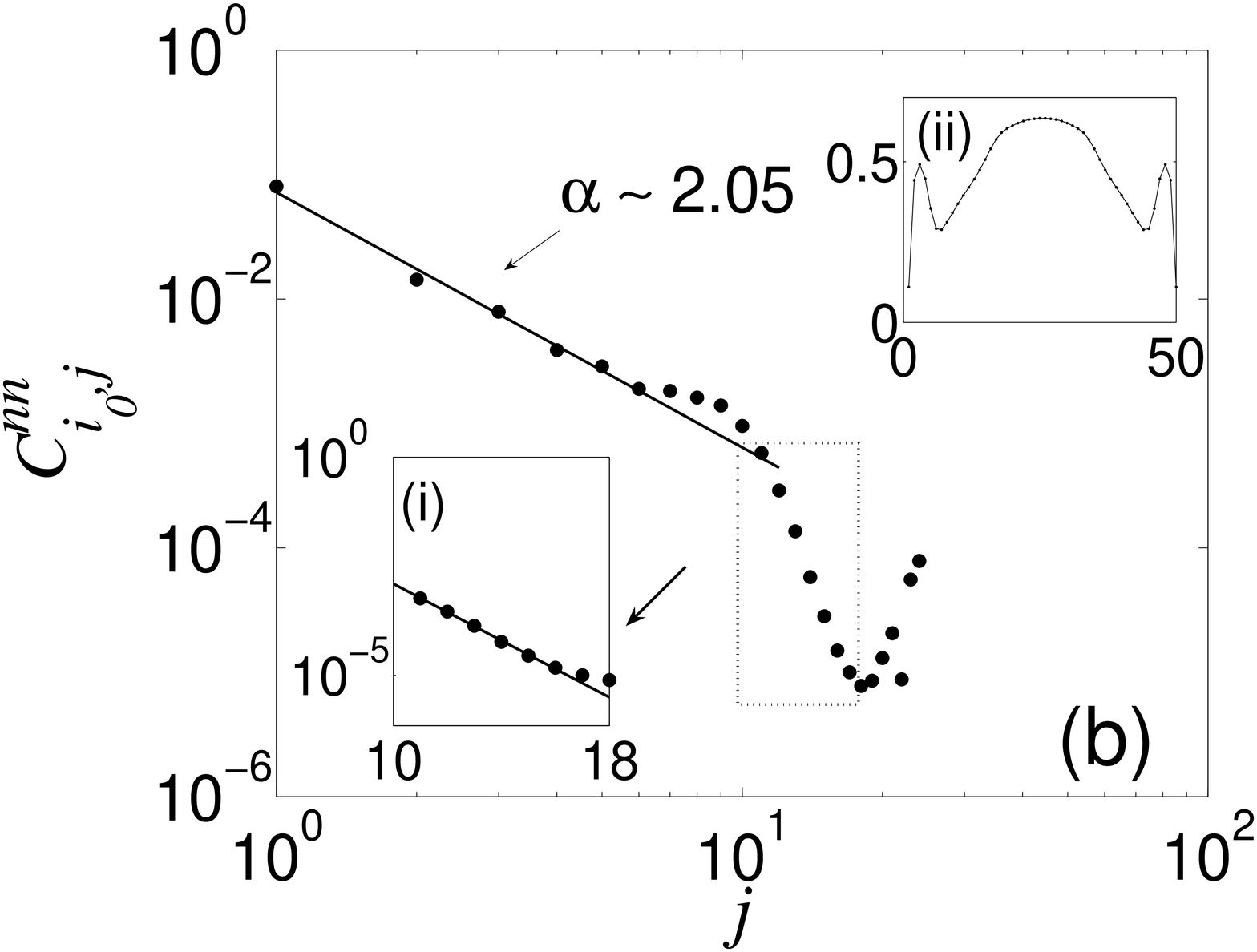}}
\caption{Correlation functions (a) $C^{aa}_{i_0,j}$ and (b) $C^{nn}_{i_0,j}$ in a linear
  Paul trap when $U/t=2$ ($N = N_\textmd{ph} = 50$). 
Both of them show the coexistence of superfluid and
  Mott-insulator phases. In (a) and (b), the insets (i) show the exponential decay in
  the region of the Mott phase, and the insets (ii) show the occupation number
  and fluctuations at the same parameters. The exponents $\alpha$ of the
  algebraic decay are also given in the figures. The dotted and solid lines
  are numerical and fitting data, respectively.}
\label{PaulCorrSFMI}
\end{figure}

Due to the localization of phonons as we increase $U/t$, 
a Mott insulator phase appears first at the sides of the ions chain,
which coexists with a superfluid core at the center. This coexistence
of the phases can be observed in the correlation functions, which show
regions of algebraic or exponential decay, as shown in Fig. \ref{PaulCorrSFMI}. 

\subsection{Mott-insulator phase}
In the commensurate case, by increasing the on-site interaction $U$, a
quantum phase transition from a superfluid to a Mott-insulator state
takes place at about $U \approx 2 t$. 
In the limit in which interaction dominates over hopping, the
ground state for a commensurate filling of $\bar{n}$ particles per site is
simply a product state of local phonon Fock states,
\begin{eqnarray}
|\psi_{MI}\rangle = \prod_{i=1}^N\frac{1}{\sqrt{\bar{n}!}}
(a^{\dagger}_i)^{\bar{n}}|0\rangle. 
\end{eqnarray}
In the Mott insulator phase
correlations decay exponentially with distance,
$C^{aa,nn}_{i,j} \propto {\textrm e}^{- |i-j|/\xi}$, where $\xi$
is the correlation length. The correlation length diverges when
approaching the quantum phase transition. 

\begin{figure}
\resizebox{1.68in}{!}{\includegraphics{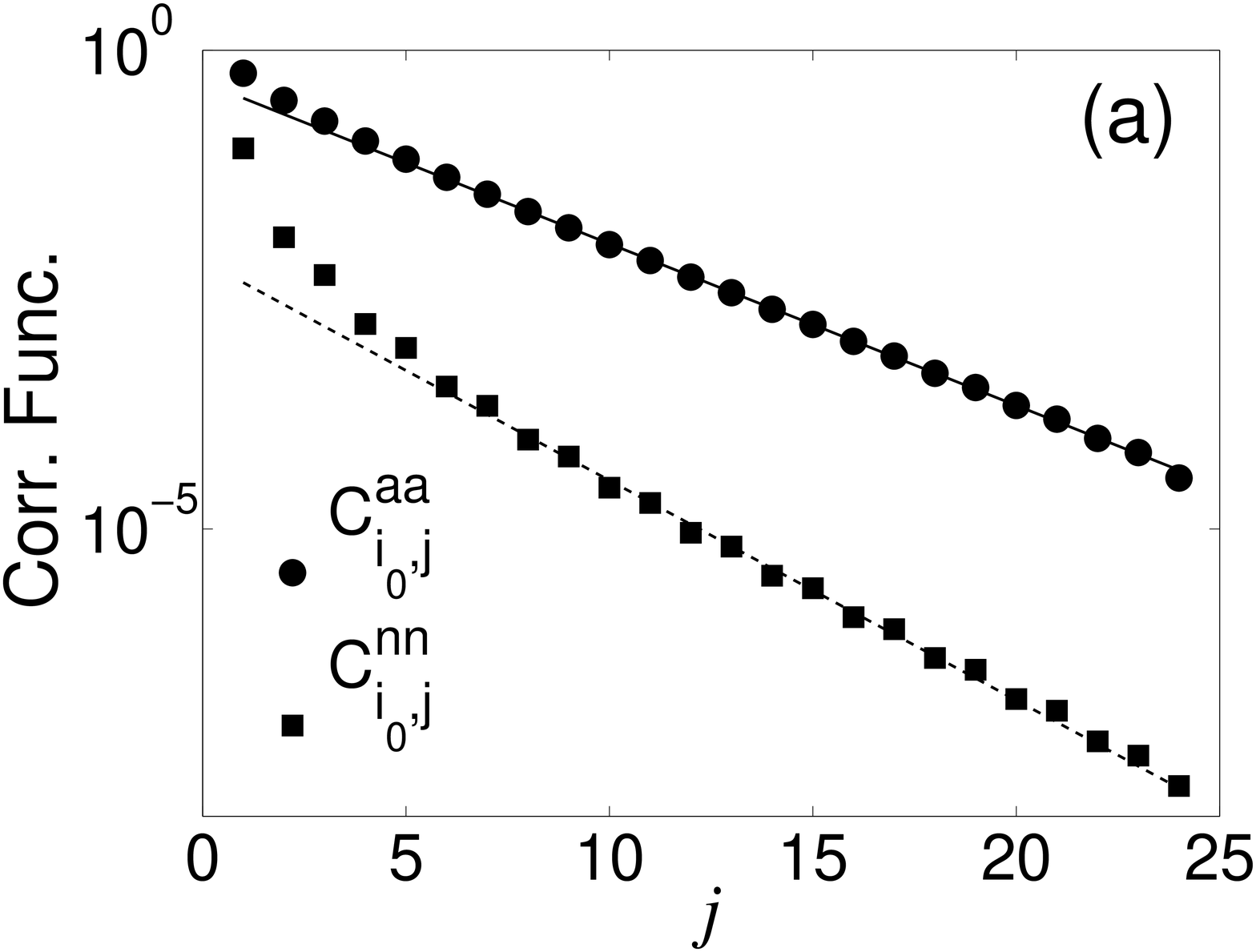}}
\resizebox{1.68in}{!}{\includegraphics{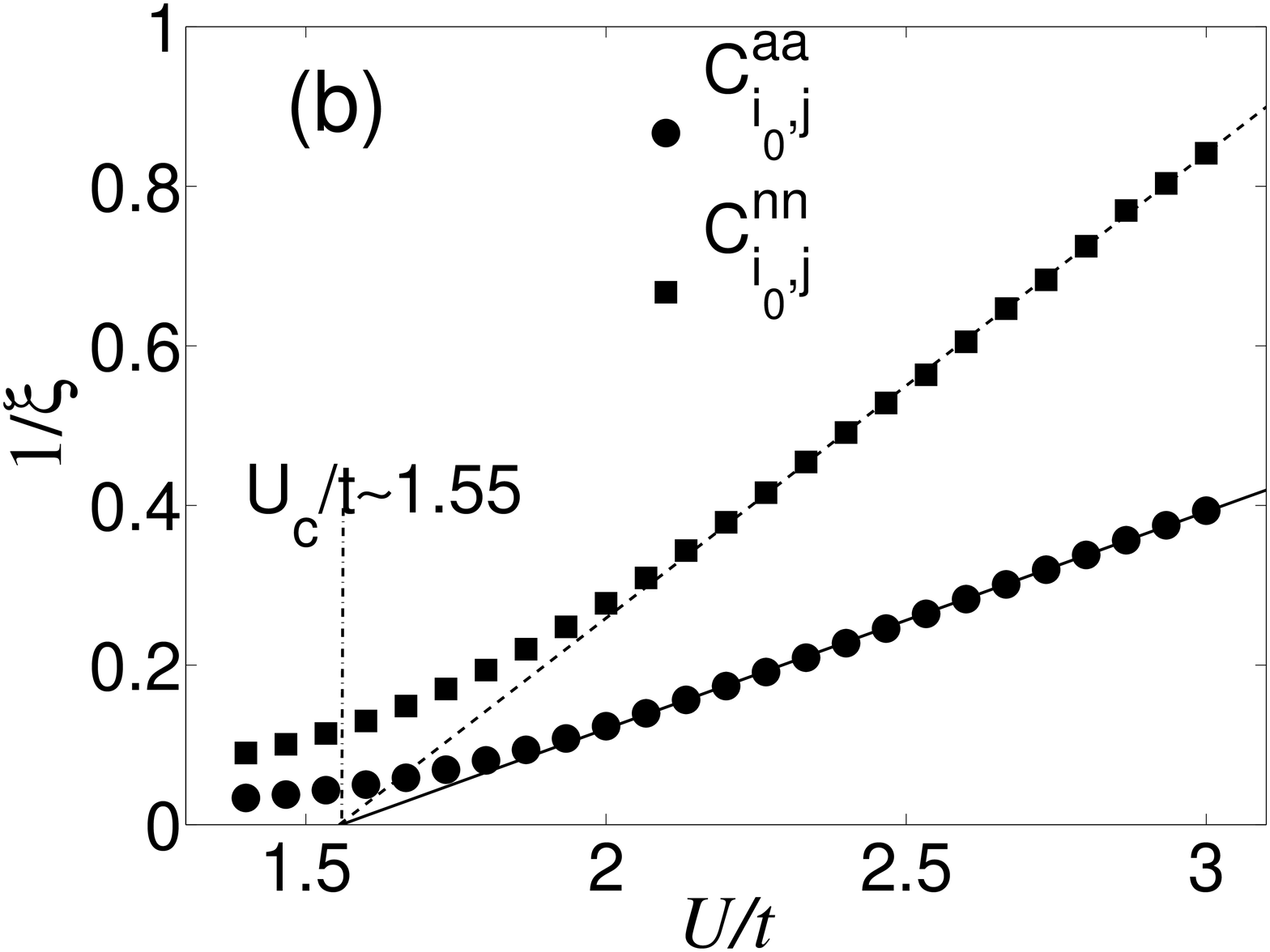}}
\caption{ (a) Correlation functions $C^{aa}_{i_0,j}$ at $U/t=3$ and
  $C^{nn}_{i_0,j}$ at $U/t=2.4$ decay exponentially with the coordinate $j$ at
  the Mott-insulator phase for an array of microtraps ($N =
  N_\textmd{ph} = 50$, $i_0 = 26$).
(b) The inverse of the
  correlation lengths in $C^{aa}_{i_0,j}$ and $C^{nn}_{i_0,j}$. The circle
  markers represent $C^{aa}_{i_0,j}$ and the square markers
  $C^{nn}_{i_0,j}$. The solid lines are the fitting data. The extrapolation in
  (b) show that the critical point in the microtraps is $U_c/t\approx 1.55$.}
\label{MicroCorrMI}
\end{figure}

{\it (1) Array of ion microtraps.} 
In Fig. \ref{MicroCorrMI} (a) we plot
the correlation functions in the phonon Mott phase 
in the case of an array of ion microtraps. 
These curves can be fitted to an exponential decay, and the correspond
correlation lengths are plotted in Fig. \ref{MicroCorrMI} (b) as a
function of the interaction strength. Due to the finite size of the
system, $\xi$ does not diverge at the critical value of $U$. 
However, the extrapolation of the curves in
the linear regime allows us to estimate the critical point, which lies at  
$U_c/t \approx 1.55$. 
This critical value is smaller than the one in 
the BHM with tunneling between nearest--neighbors only, $U^{nn}_c/t\approx 2$. 
The condition $U_c < U_c^{nn}$ is 
due to the frustration induced by hopping between
next--nearest--neighbors, which makes the superfluid phase more
unstable against the effect of on--site interactions.

\begin{figure}
\resizebox{1.68in}{!}{\includegraphics{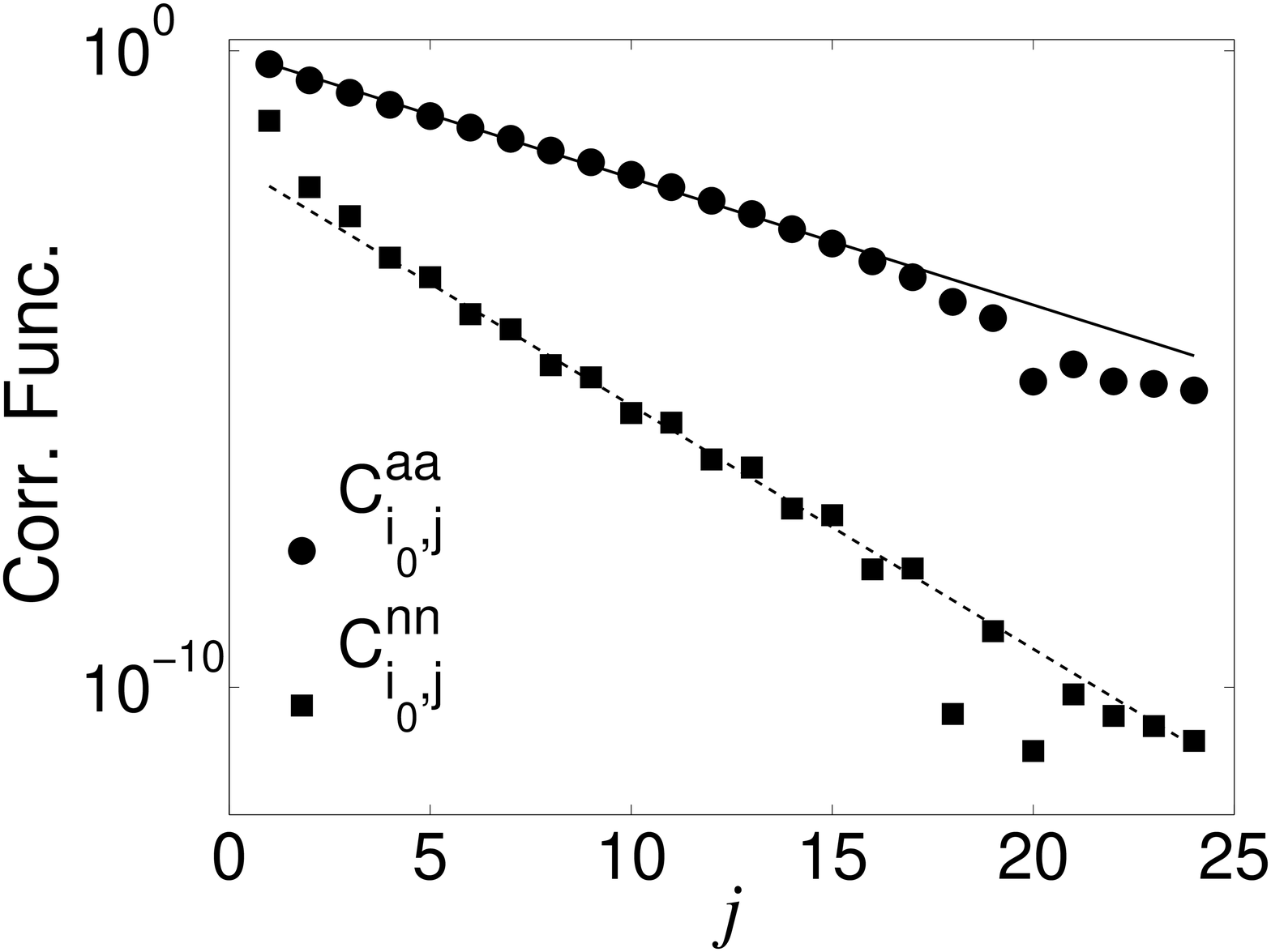}}
\caption{The exponential decay correlations $C^{aa}_{i_0,j}$ and $C^{nn}_{i_0,j}$ at
  the Mott phase $U/t=2.8$ for a linear Paul trap ($N = N_\textmd{ph}
  = 50$, $i_0 = 26$).}
\label{PaulCorrMI}
\end{figure}

{\it (2) Linear ion trap.} 
In the case of a linear trap, the behavior of  spatial correlations is similar, see
Fig. \ref{PaulCorrMI}.  
$C^{nn}_{i,j}$ is difficult to fit due to the few
points with exponential decay, therefore we only plot the correlation length
corresponding to $C^{bb}_{i,j}$. 
Due to the effective phonon trapping potential, the Mott insulator and
superfluid phases coexist in a range of values of $U$
(Fig. \ref{PaulCorrSFMI}). For this reason, in the case of a linear
Paul trap, one cannot follow the extrapolation procedure of Fig. 
\ref{MicroCorrMI} to find a critical value of the interaction.

\begin{figure}
\resizebox{3.0in}{!}{\includegraphics{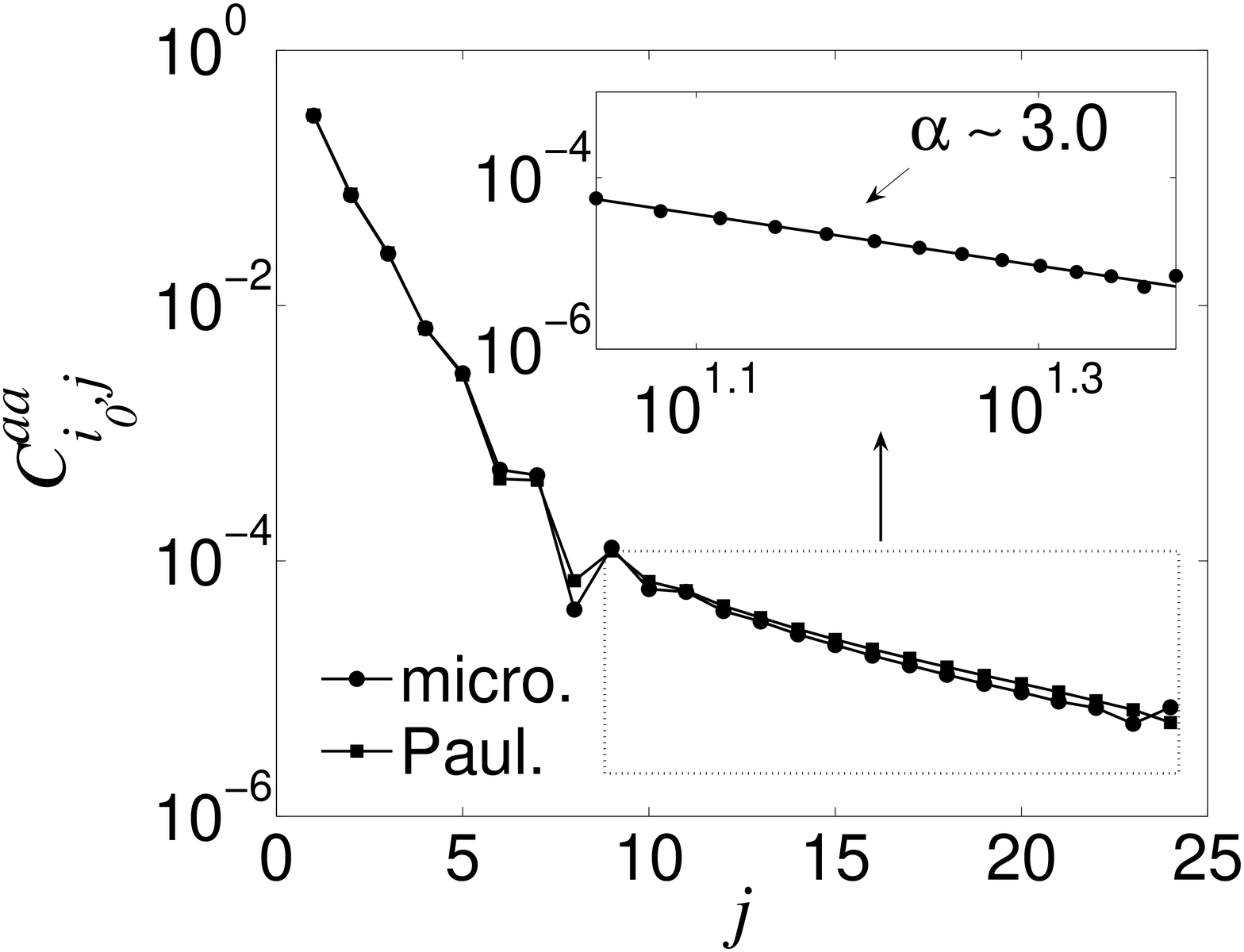}}
\caption{The correlation $C^{aa}_{i_0,j}$ at the Mott phase
  $U/t=6$ in an array of microtraps (circle markers) and a linear Paul trap
  (square markers), respectively. $N = N_\textmd{ph} = 50$, $i_0 =
  26$.
 The inset shows the power-law decay
  with exponent $\alpha\approx 3$ for the microtraps only.}
\label{MottPower}
\end{figure}

Finally, in the Mott-insulator phase the long-range hopping terms 
which decay like
$1/|z^0_i - z^0_j|^3$ play
a major role, since they induce a peculiar long-range correlation in this
phase. In Fig. \ref{MottPower}, we show that $C^{aa}_{i,j}$ indeed
also behaves like
$1/|z^0_i - z^0_j|^{3}$ at long distances. The existence of power--law decay in
correlation functions of non--critical systems due to long--range
interactions was also observed in the case of spin models in 
trapped ions, see the discussion in Ref. \cite{DengPorrasCirac.spin}.

\subsection{Tonks-gas phase}
We turn now to the incommensurate filling case, where 
in the limit $U \gg t$ the system forms a Tonks-Girardeau gas, which
can be described in terms of effective free fermions.
A Tonks-Girardeau gas has recently been realized in
an experiment with ultracold bosons in an optical lattice
\cite{Paredes.Tonks}.

\begin{figure}
\resizebox{1.68in}{!}{\includegraphics{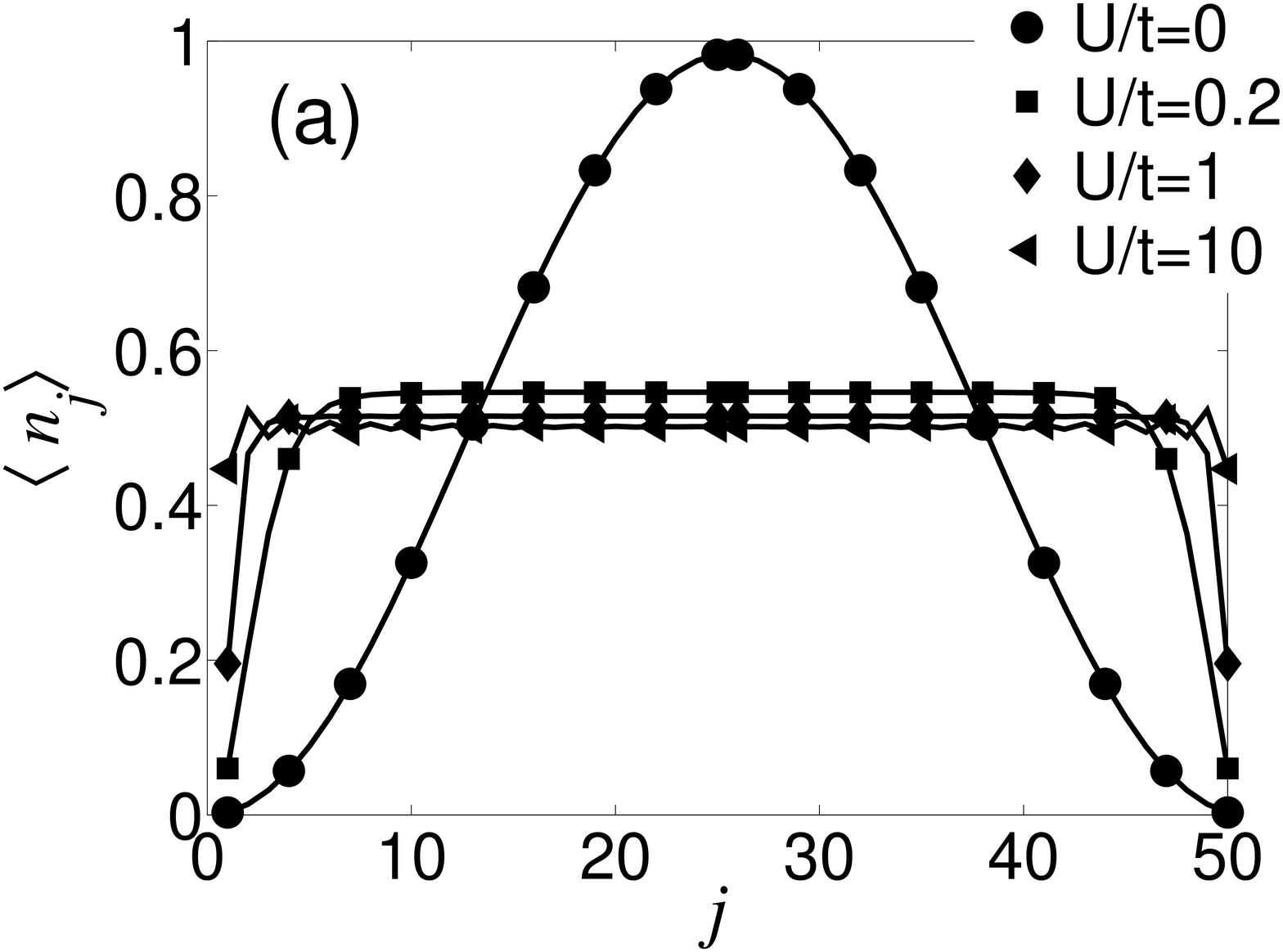}}
\resizebox{1.68in}{!}{\includegraphics{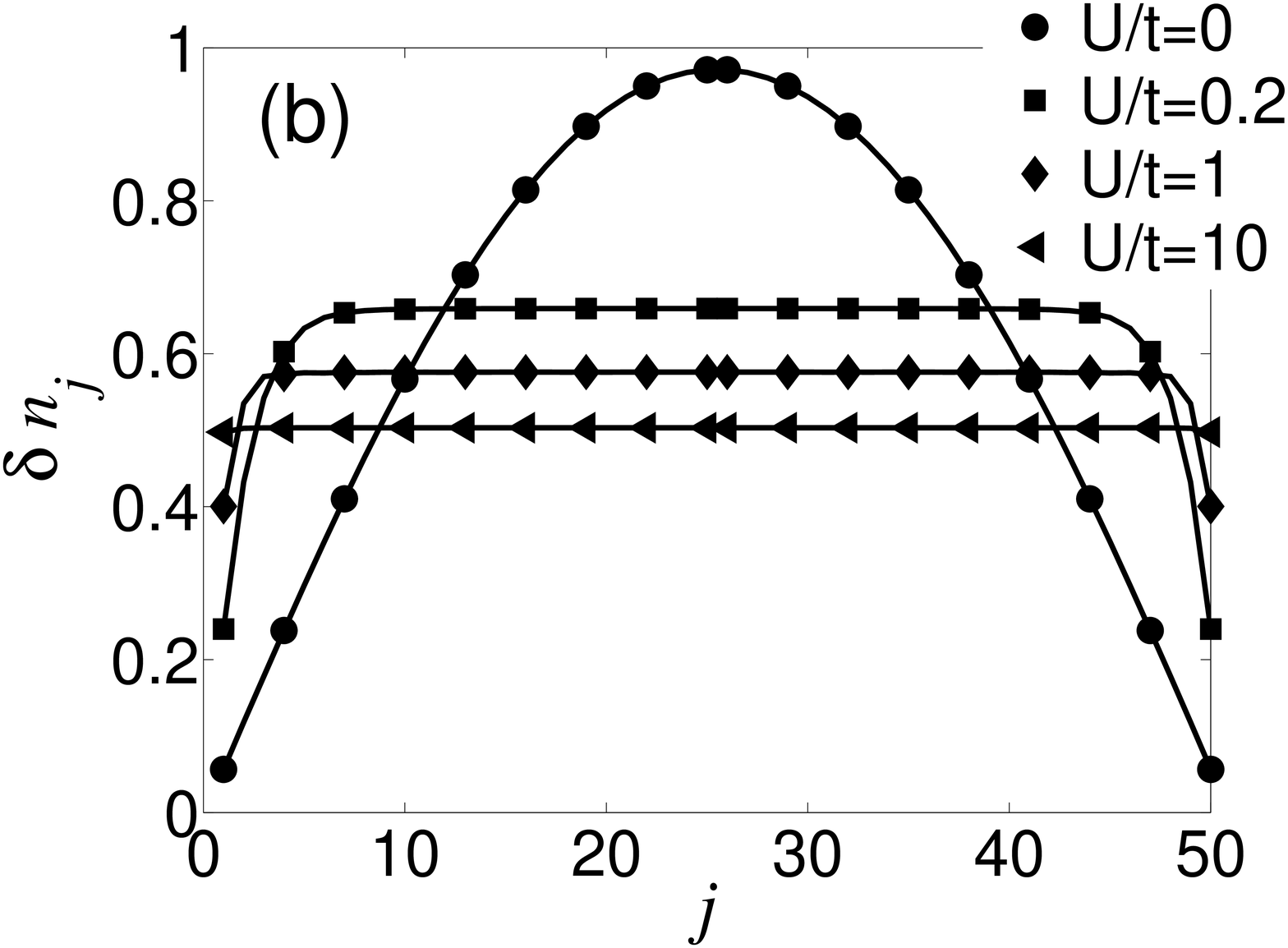}}
\resizebox{1.68in}{!}{\includegraphics{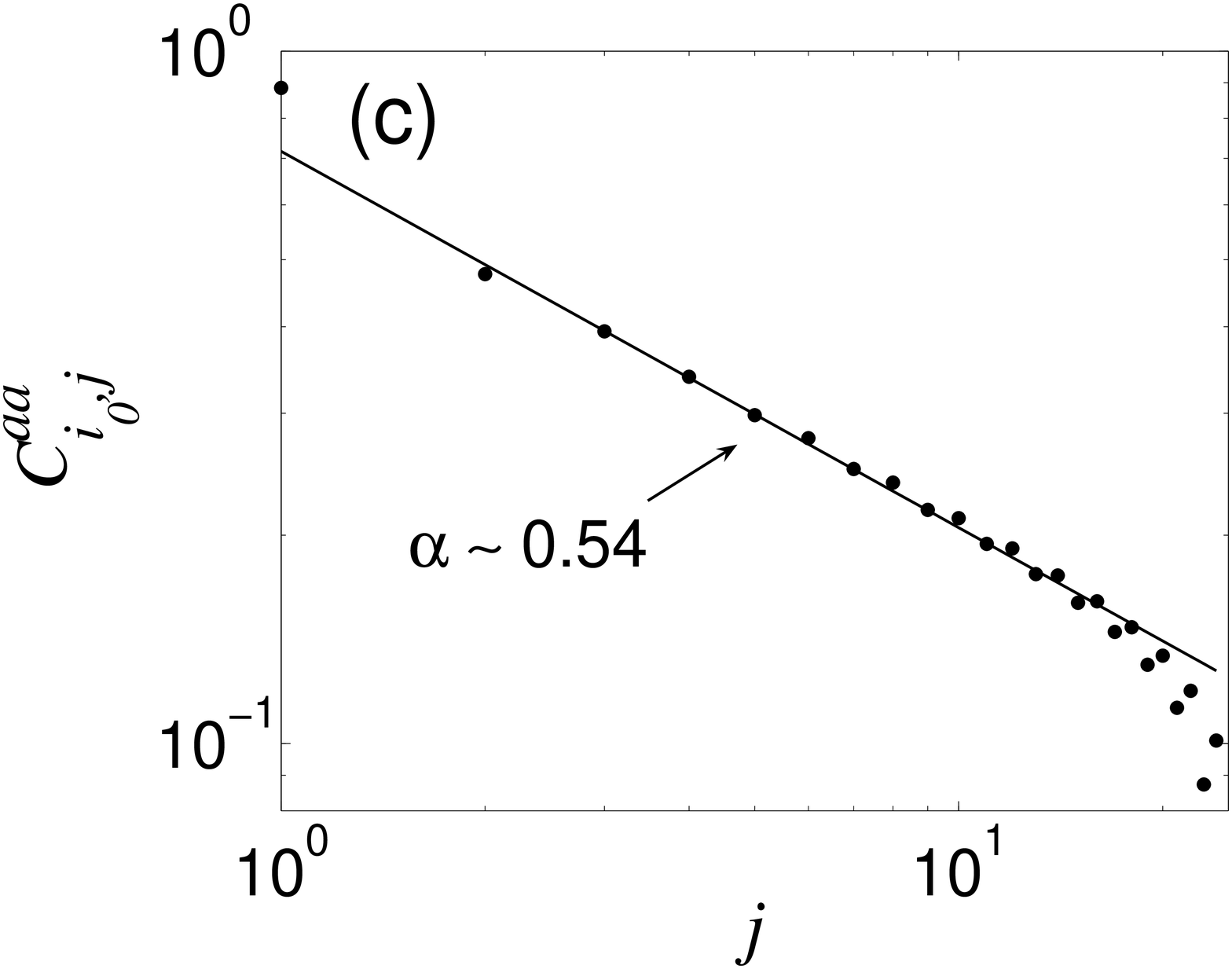}}
\resizebox{1.68in}{!}{\includegraphics{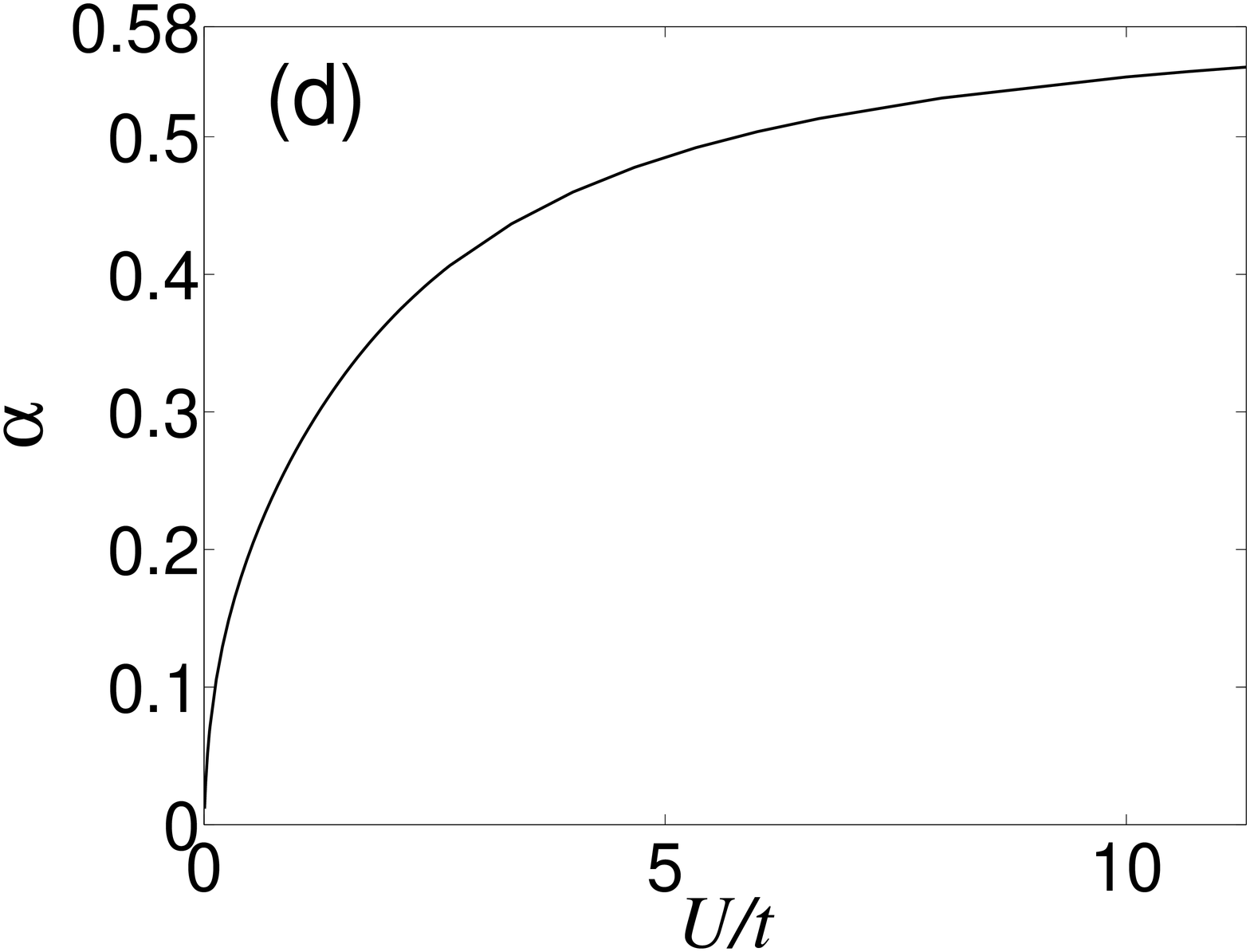}}
\caption{(a) Densities of phonons in an array of microtraps with $N=50$, $N_{ph}=25$. (b)
  Fluctuations at the same conditions. 
(c) The correlation function $C^{aa}_{i_0,j}$ ($i_0 = 26$) at Tonks-gas
  phase $U/t=10$ with exponent $\alpha\approx 0.54$, where the dotted and
  solid lines are numerical and fitting data, respectively. (d) Evolution of exponent
  $\alpha$ of $C^{aa}_{i_0,j}$ with the interaction
  $U/t$, which would approach $\alpha\approx 0.58$.}
\label{MicroIncomm}
\end{figure}

{\it (1) Array of ion microtraps.}
We have studied numerically the Tonks--Girardeau regime in the case of
phonons in ion traps, starting with the case of an array of ion microtraps
with $N = 50$ sites and $N_{ph} = 25$ phonons, that is, $1/2$ filling. 
The density of phonons evolves from a
superfluid to a Tonks-gas profile when increasing the interaction, and at
the end it approaches a constant value of $1/2$. 
Correlation functions decay algebraically, with an exponent 
that approaches $\alpha \approx 0.58$ for large interactions (see
Fig. \ref{MicroIncomm}).
Note that $\alpha$ deviates from $1/2$, which is the value that
corresponds to a Tonks gas with nearest--neighbor tunneling only. The
deviation can be explained by the mapping from the BHM
model (\ref{BHM}) with $U \gg t$ to an XY model with antiferromagnetic 
interactions of the form $J_{i,j} = J/|i-j|^3$. The long--range terms
in the antiferromagnetic interaction induces a change in the exponent
of the correlation functions, as shown with the numerical calculations
of our previous work on spin models in ion traps \cite{DengPorrasCirac.spin}. 
%

\begin{figure}
\resizebox{1.68in}{!}{\includegraphics{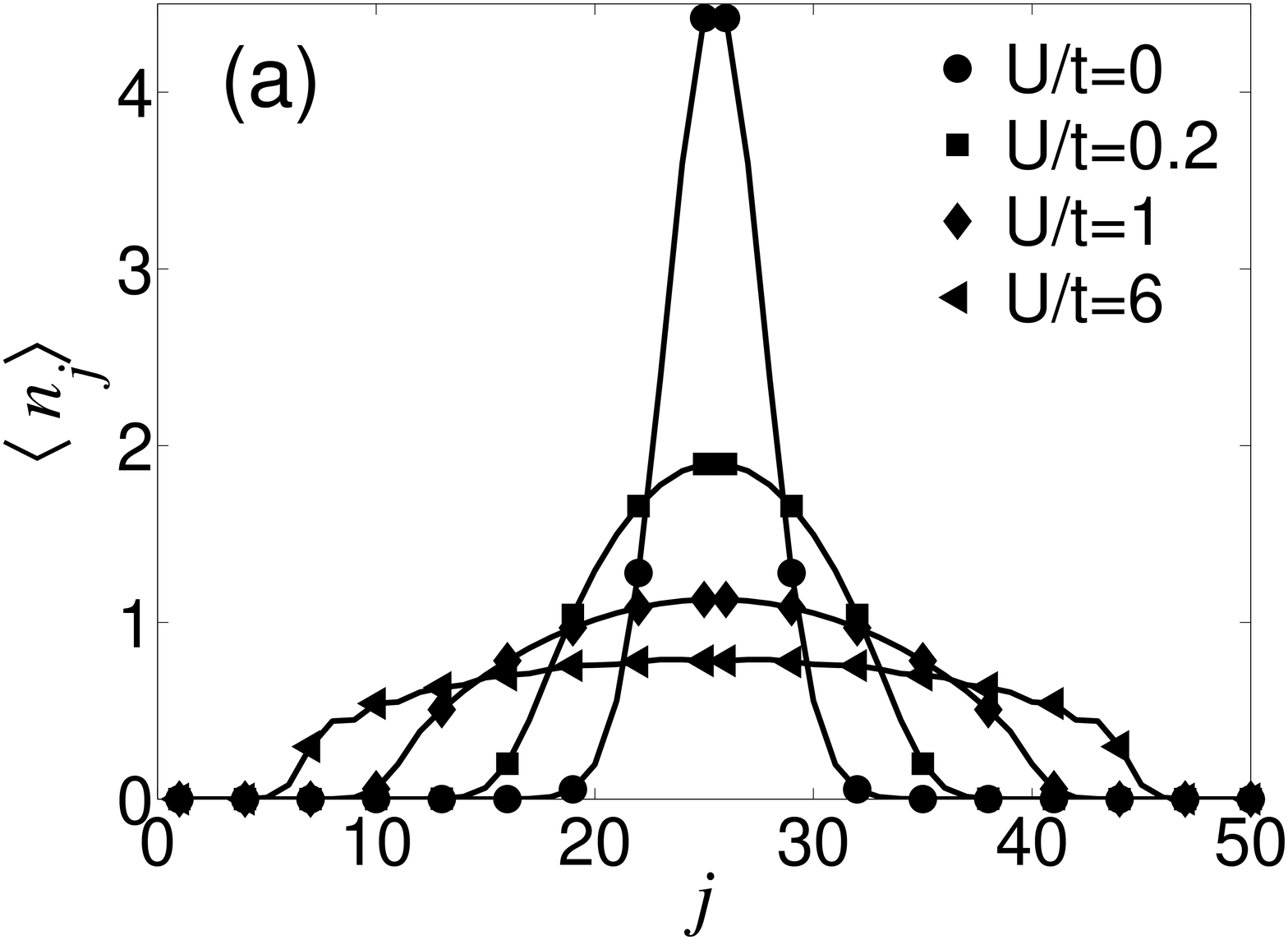}}
\resizebox{1.68in}{!}{\includegraphics{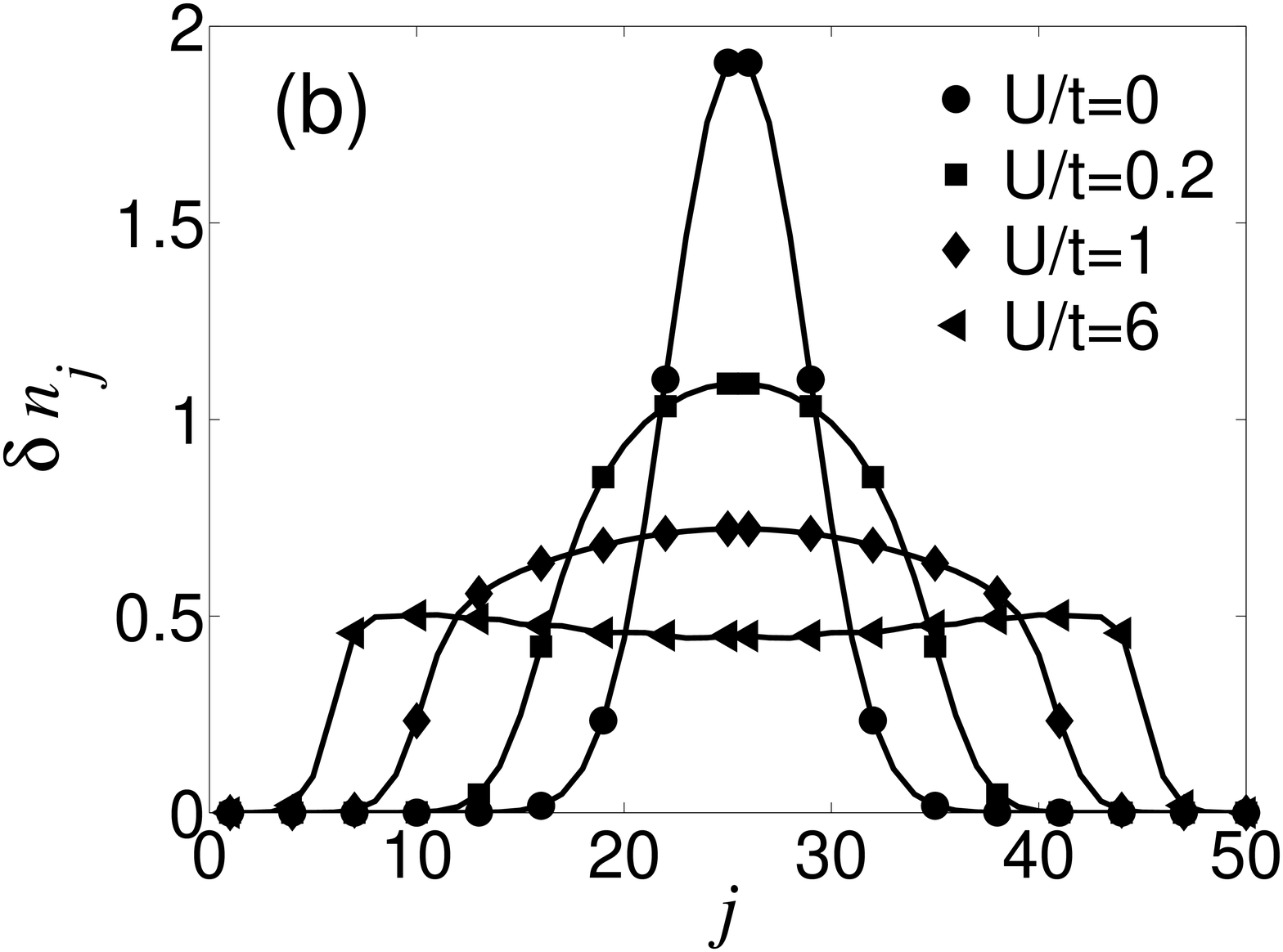}}
\resizebox{1.68in}{!}{\includegraphics{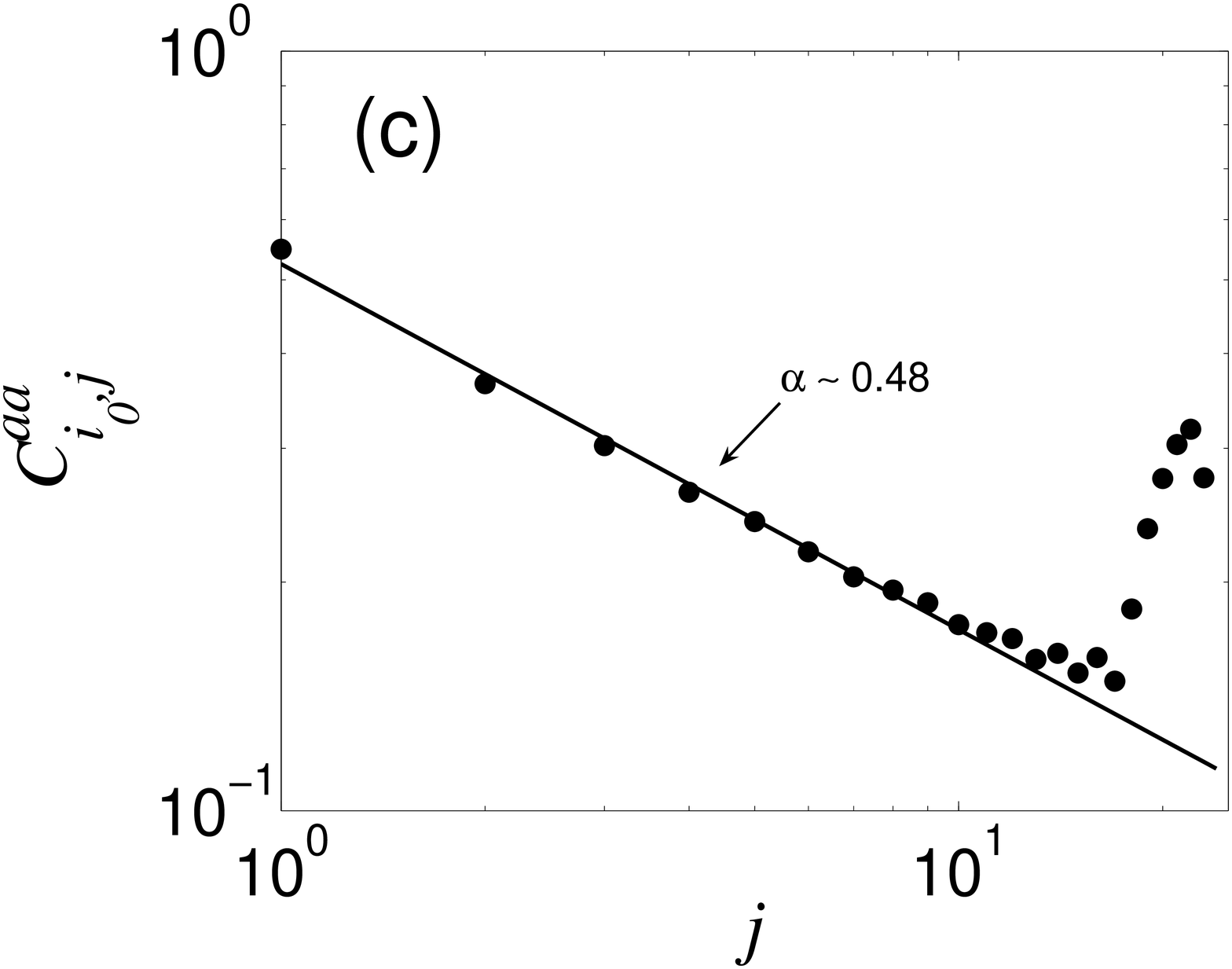}}
\resizebox{1.68in}{!}{\includegraphics{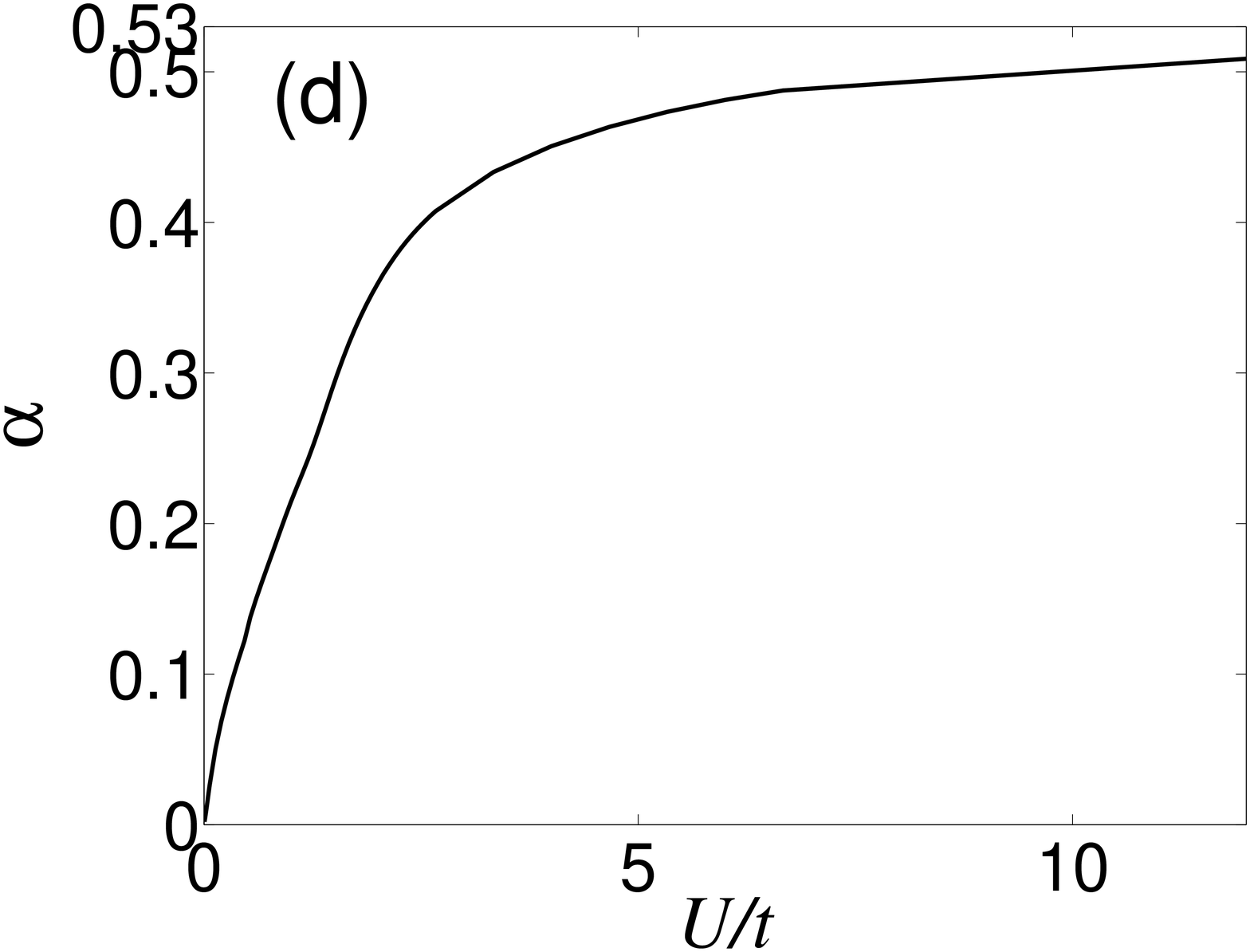}}
\caption{(a) and (b) show densities and fluctuations of phonons, respectively, in a linear
Paul trap with $N=50$ and $N_{ph}=50$; (c) The correlation $C^{aa}_{i_0,j}$ at
the Tonks-gas phase $U/t=6$, whose exponent is $\alpha\approx 0.48$; (d)
Evolution of the exponents $\alpha$ of $C^{aa}_{i_0,j}$, approaching
$\alpha\approx 0.53$. In (c) the dotted and solid lines represent numerical
and fitting data, respectively.}
\label{PaulIncomm}
\end{figure}

{\it (2) Linear ion trap.}
We study now the case of phonons in a linear Paul
trap under the same conditions, see Fig. \ref{PaulIncomm}. Correlation functions decay
algebraically, with an exponent that is extrapolated to 
$\alpha = 0.53$ in the limit of strong interactions.  This result can
also be explained by the mapping to the $XY$ model, and coincides with
the result found in \cite{DengPorrasCirac.spin}.

\begin{figure}
\resizebox{3.0in}{!}{\includegraphics{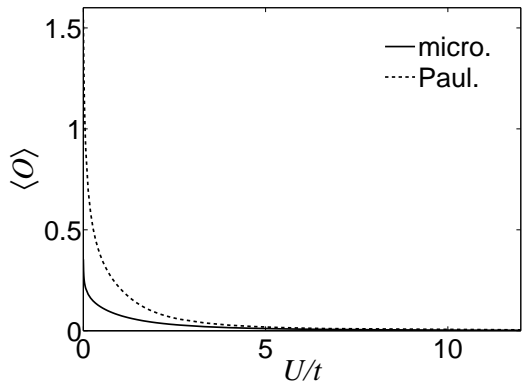}}
\caption{The evolution of the parameter $\langle O\rangle$ with the
  interaction $U/t$ in an array of microtraps (solid line) and a linear Paul
  trap (dashed line). $N =  50$,  $N_\textmd{ph} = 25$.}
\label{MPIncommEvol}
\end{figure}

In order to test if the system is really in the Tonks--gas phase,
we introduce the quantity 
$\langle O \rangle =\langle\sum_i n_i(n_i -1)\rangle/N$, which
measures the probability of phonon occupancies larger than one. 
In the Tonks--gas regime $\langle O \rangle \sim 0$.
The parameter $\langle O\rangle$ as a function of the
interaction $U$ is plotted in Fig. \ref{MPIncommEvol}, showing the
continuous evolution into the Tonks--gas regime.

Our results are consistent with the behaviors observed in
optical lattices \cite{Paredes.Tonks}. The numerical analysis shows
that phonons in ion traps are also a good candidate like atoms in
optical lattices for studying Tonks gases.

\section{Bose-Hubbard model with Attractive interactions: $U<0$}
\label{section.attractive.interactions}
The BHM with attractive interactions in optical lattices has been
the focus of recent theoretical studies \cite{JackYamashita}. 
The sign of phonon--phonon interactions in trapped
ions can be made negative simply by changing the relative position of the
standing--wave relative to the ion chain. 
For a qualitative understanding of this model, it is useful to
consider a Bose--Einstein condensate in a double well potential \cite{JackYamashita,
  SteelCollett}. 
In a symmetric potential in the absence of tunneling, energy 
is decreased when bosons accumulate in
one of the wells. When tunneling is switched on, the ground state of
the system is a linear superposition of states with all the bosons
placed in one of the wells, showing large phonon number fluctuations.

\begin{figure}
\resizebox{1.68in}{!}{\includegraphics{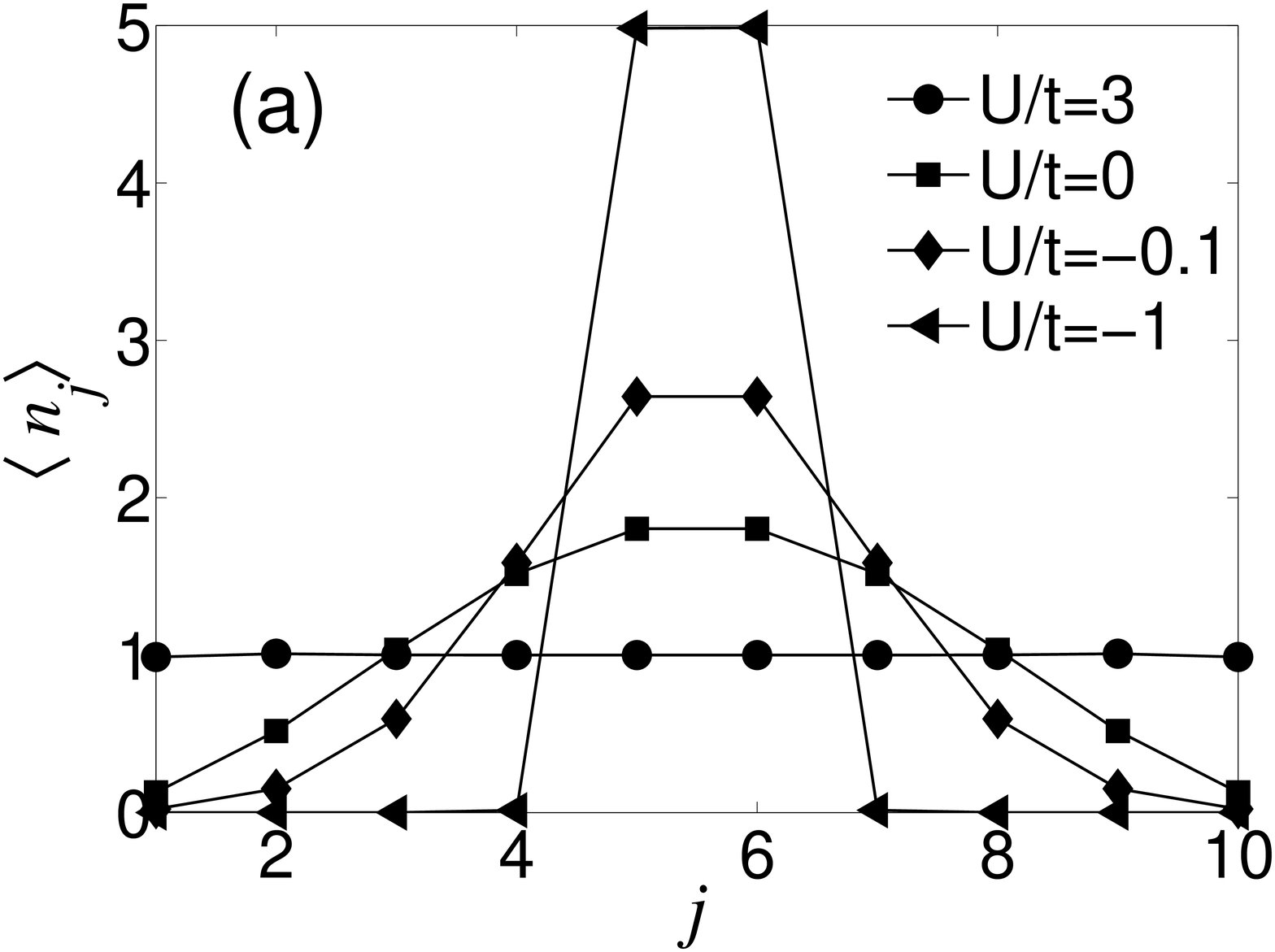}}
\resizebox{1.68in}{!}{\includegraphics{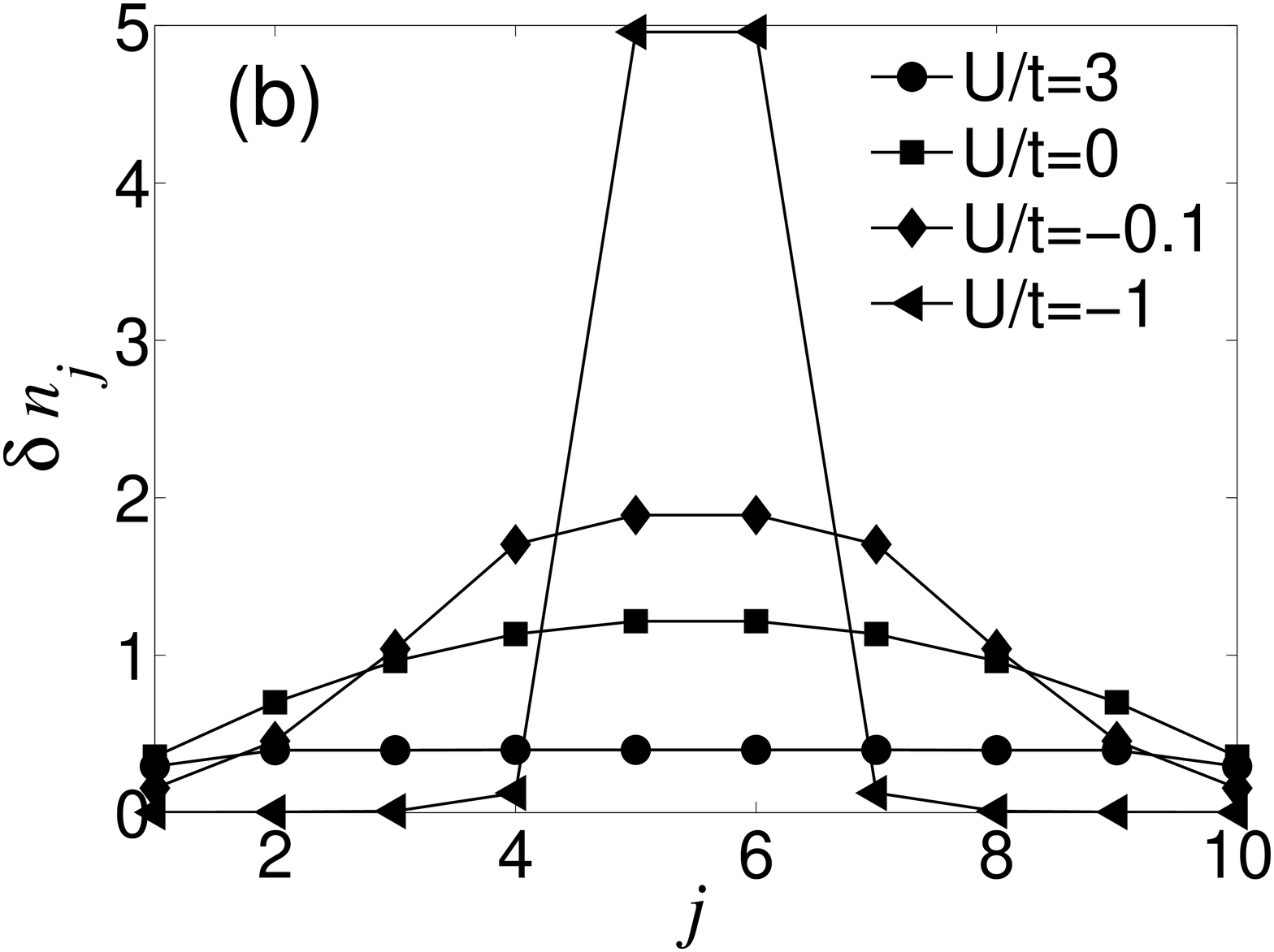}}
\resizebox{1.68in}{!}{\includegraphics{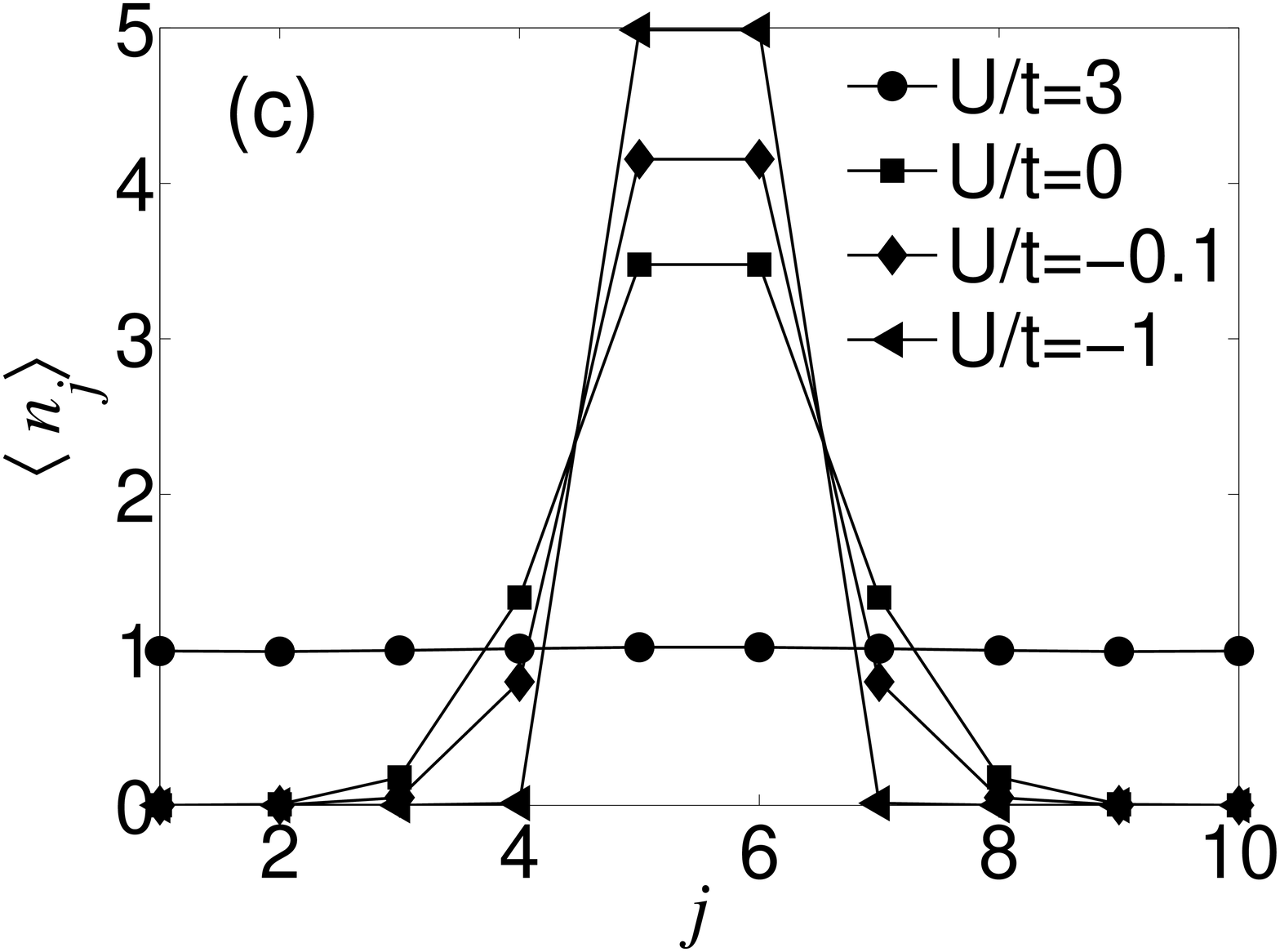}}
\resizebox{1.68in}{!}{\includegraphics{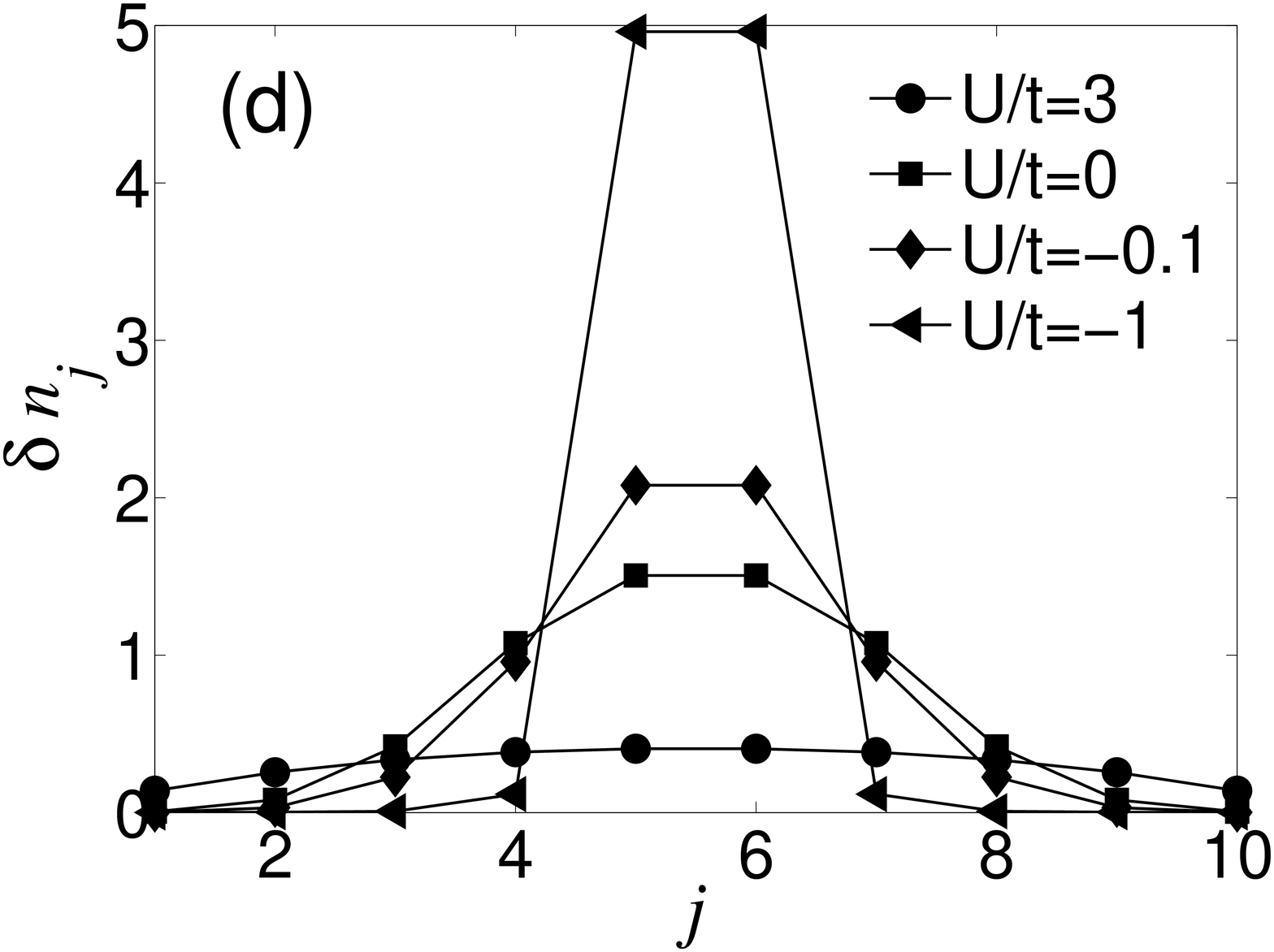}}
\caption{(a) Density of phonons and (b) phonon number fluctuations 
in an array of microtraps with $N=10$, $N_{ph}=10$, and negative
on--site interactions . 
(c) Density of phonons and (d) phonon number fluctuations in a
linear Paul trap under the same conditions.}
\label{Negative}
\end{figure}

The increase of phonon number fluctuations in our model when we
switch on a negative interaction is also shown in our numerical
calculations. In Fig. \ref{Negative}, the density at the center of the
ion chain and its fluctuations 
increase with the magnitude
of the interaction for $N=10$ ions and $N_{ph}=10$ phonons with open boundary
conditions. 
Due to the open boundary condition and the symmetry of the
potential, the phonons tend to collect themselves on one of the two sites at the center when
increasing $|U|$ with an even number of sites. The ground state  is
then a superposition of $N_{ph}$ phonons on site $N/2$ and $N_{ph}$ phonons on site
$N/2+1$. 

\begin{figure}
\resizebox{3.0in}{!}{\includegraphics{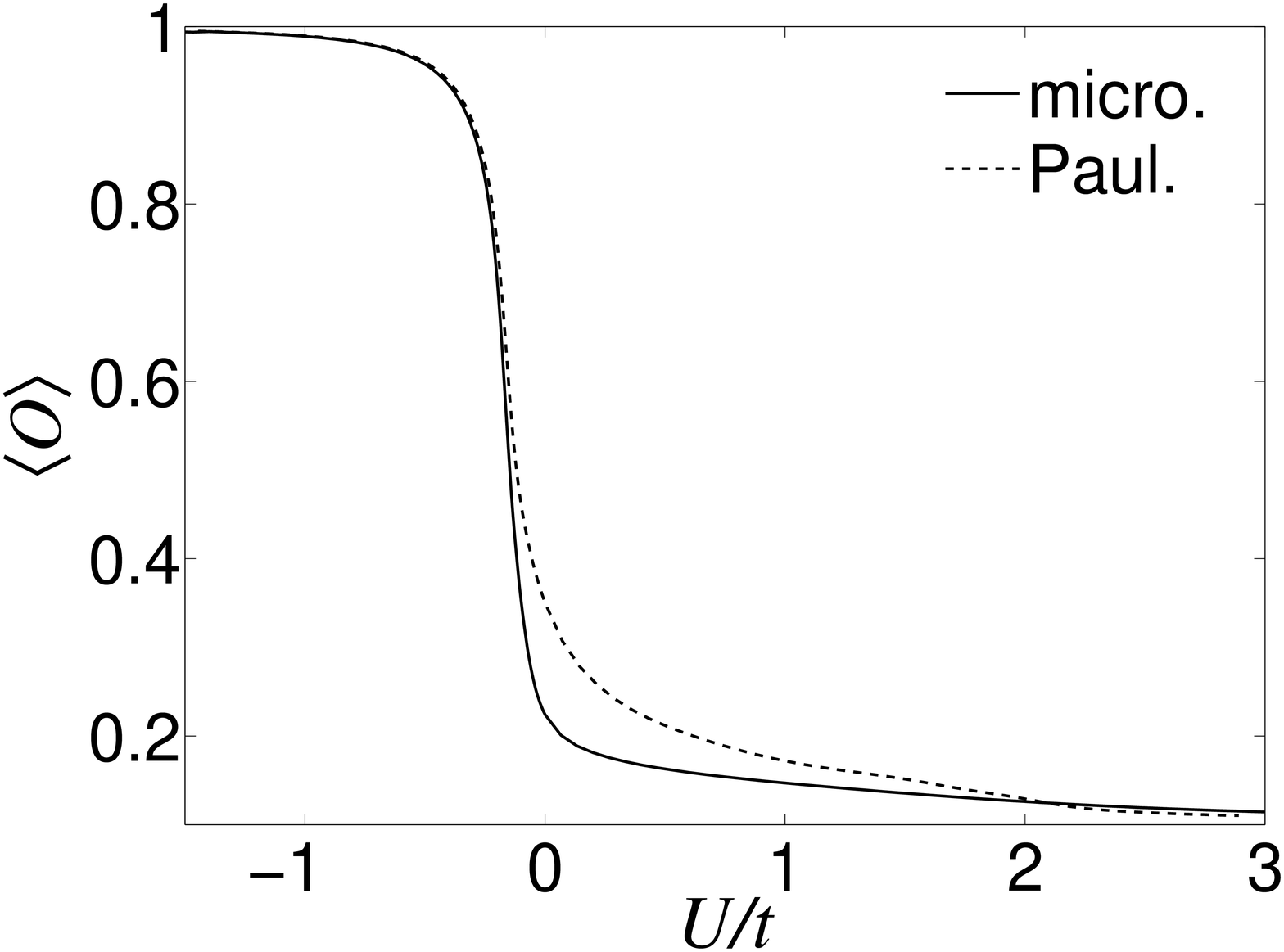}}

\caption{Evolution of the order parameter $\langle O \rangle$ with the
  ratio $U/t$ for an array of ion microtraps (solid line) and a linear Paul
  trap (dashed line). $N = N_\textmd{ph} = 10$.}
\label{NegativeEvol}
\end{figure}

When $|U|$ is large enough, our numerical calculations yield a ground
state with all the phonons in a single ion, such that the spatial
symmetry of the problem is broken. This effect is an artifact of the
DMRG calculation, due to the small energy difference between the exact
ground state of the system and the one which breaks the spatial
symmetry. Thus, in order to study properly the phonon phases with
negative interaction, it is convenient to define the following order parameter,
whose value is independent on the breaking of the spatial symmetry in
the problem:
\begin{eqnarray}
\langle O\rangle=\frac{1}{N^2}\langle\sum_{j}(a^{\dagger}_j a_j)^2\rangle.
\end{eqnarray}
In Fig. \ref{NegativeEvol} we show the evolution of this
quantity, which shows a sudden increase for negative interactions.  

\section{Bose-Hubbard model with site-dependent interactions}
\label{section.site.dependent.interactions}
Phonons in trapped ions have a higher controllability than
ultracold neutral atoms in optical lattices, due to the possibility of
individual addressing.
In particular, on--site interactions can be induced in such a way that
they depend on the ion position. In this section we present a model
which shows how this possibility can be exploited for
the engineering of quantum phases.  

Let us consider repulsive on-site interactions which vary over the ion
chain in the following way:
\begin{eqnarray}
U_i =  U_\textmd{odd}  =  U, \hspace{0.6cm} && \textmd{i  odd}  ,    \nonumber \\
U_i =  U_\textmd{even} = 2U, \hspace{0.6cm} && \textmd{i  even} ,    
\label{alternating.interactions}
\end{eqnarray}
We focus on the case with filling factor 2, $N_\textmd{ph} = 2 N$, 
in the regime where interactions dominate over
tunneling, $U / t \gg 1$. 
In the limit $t = 0$, the ground state of this model is highly degenerate. 
For instance, in a chain with two sites
the ground state manifold in the Fock basis spans the states 
$| 2, 2 \rangle$
and
$| 3, 1 \rangle$. 
In a chain with even number $N$ of sites and $N_\textmd{ph} = 2 N$,
  the ground state degeneracy is
$\left( \begin{array}{c} N \\ N/2 \end{array} \right)$ .

\begin{figure}
\resizebox{2.5in}{!}{\includegraphics{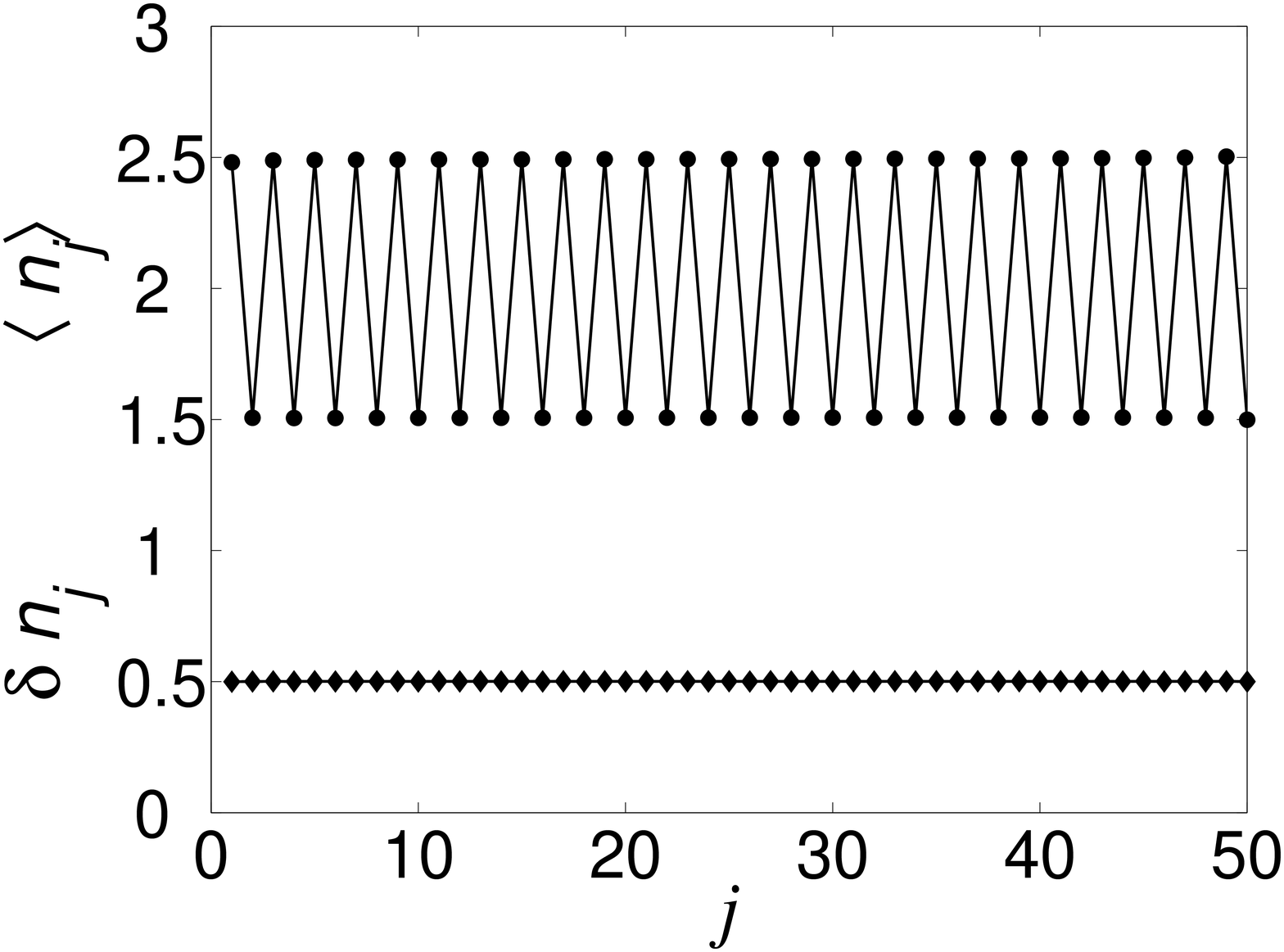}}
\caption{The phonon density $\langle n_j \rangle$ and phonon number
  fluctuations, $\delta n_j$, in an array of ion microtraps, with the
  on--site interactions defined by
  Eq. (\ref{alternating.interactions}),
$U/t=40$, $N=50$, and $N_{ph}=2N=100$.}
\label{DegenerateDensity}
\end{figure}

Our DMRG algorithm allows us to calculate the density and fluctuations
in the phonon number, which are shown in
Fig. \ref{DegenerateDensity}, for the case where the interactions
defined by Eq. (\ref{alternating.interactions}) are induced on an
array of ion microtraps. In the ground state of the chain, the number
of phonons fluctuates between $| 2 \rangle$ and $| 3 \rangle$, and 
$| 2  \rangle$ and $| 1 \rangle$, in odd and even sites,
respectively. 

This model can be understood in the hard--core boson limit by
introducing a spin representation, which is valid near
filling factor 2. At each site in the ion chain, we define a two level
system by means of the following rule:
\begin{eqnarray}
| \bar{0} \rangle_i = | 2 \rangle_i, \hspace{0.2cm}
| \bar{1} \rangle_i = | 3 \rangle_i, \hspace{0.1cm}  & & \textmd{i  odd}, 
\nonumber \\
| \bar{0} \rangle_i = | 1 \rangle_i, \hspace{0.2cm} 
| \bar{1} \rangle_i = | 2 \rangle_i, \hspace{0.1cm}  & & \textmd{i  even} .
\label{effective.spin.definition}
\end{eqnarray}
Where $| \bar{0} \rangle$ and $| \bar{1} \rangle$ are the two levels
which define the spin states of the spin representation of the
hard--core bosons.  Spin and  phonon annihilation
operators satisfy:
\begin{eqnarray}
\sigma^{+}_{i} &=& \frac{1}{\sqrt{3}} a^{\dagger}_i,
\hspace{0.2cm} \textmd{i odd}, 
\nonumber\\
\sigma^{+}_{i} &=& \frac{1}{\sqrt{2}} a^{\dagger}_i,
\hspace{0.2cm} \textmd{i even}.
\label{spin.operators.XY}
\end{eqnarray}
where the equality is understood to hold within the ground state
manifold. In terms of this operators, the Hamiltonian of the system is described by
an XY model (for simplicity we consider here the nearest-neighbor case):
\begin{eqnarray}
H = {\tilde J} \sum_i(\sigma^{+}_i \sigma^{-}_{i+1} + h.c.),
\label{XY}
\end{eqnarray}
with ${\tilde J} = \sqrt{6} t$. 
Under the condition $N_{\textmd{ph}}/N = 2$, the ground state of our
hard--core boson Hamiltonian corresponds to the solution of the $XY$
model (\ref{XY}) with the constraint $\sum_i \sigma^z_i = 0$.

\begin{figure}
\resizebox{2.5in}{!}{\includegraphics{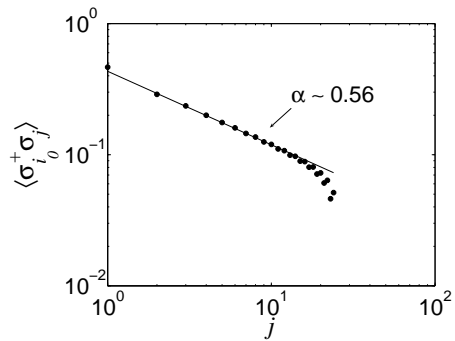}}
\caption{
The correlation function 
$\langle \sigma^{+}_{i_0} \sigma_{j}\rangle$ defined by (\ref{effective.spin.definition}).
$i_0=26$, $N = 50$, $N_\textmd{ph} = 100$, $U/t = 40$.}
\label{DegenerateCorr}
\end{figure}

Spin-spin correlation functions are related to the bosonic
correlations functions of the BHM by the relation
Eq. (\ref{spin.operators.XY}). 
In Fig. \ref{DegenerateCorr} we plot $\langle \sigma^+_{i} \sigma_j \rangle$
calculated by means of correlation functions of hard--bosons. 
This correlation function shows an algebraic decay for short
distances, which is spoiled for long separations
between ions due to boundary effects. 
The exponent $\alpha\sim 0.56$, differs from the one that we
expect from the mapping to the XY model, that is, $\alpha = 0.5$, due to the
effect of further than nearest--neighbor interactions terms, which we have
neglected.

\begin{figure}
\resizebox{2.5in}{!}{\includegraphics{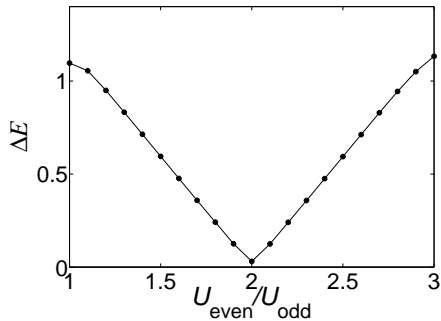}}
\caption{The energy gap $\Delta E$ between the ground state and the first excited state
  as a function of $U_\textmd{even}/U_\textmd{odd}$. $U_\textmd{odd}/t = 40$. 
For simplicity, we consider here a small system of ions $N=6$.}
\label{EnergyGap}
\end{figure}

Another interesting possibility is to study the phase diagram of the
BHM with filling factor 2, and alternating interactions, beyond the $XY$ point 
$U_\textmd{even} = 2 U_{\textmd{odd}}$. If we change the ratio 
$U_\textmd{even}/U_{\textmd{odd}}$ in the vicinity of this point, 
the ground state degeneracy with zero tunneling is lifted. The
system is in a Mott insulator phase, with constant phonon density 
if $U_\textmd{even} > 2 U_{\textmd{odd}}$, 
$|2, 2, 2, \dots \rangle$, or alternating occupation numbers if 
$U_\textmd{even} < 2 U_{\textmd{odd}}$ , $|3, 1, 3, 1, \dots \rangle$.
The phase diagram as a function of the ratio $U_\textmd{even} /
U_{\textmd{odd}}$, shows two gaped regions, separated by a single
critical point, which corresponds to the $XY$ limit studied
above. This is shown in Fig. \ref{EnergyGap}, where we calculate the
energy gap $\Delta E$ for small chains ($N = 6$).

To generalize, the same also takes place with other distributions of
the on--site interactions, whenever the number of sites where phonons interact
with $U$ is the same as the number of sites with $2 U$ interaction.
For example, the chain can be divided in two regions, left and right,
such that the interactions depend on the site in the way
$U,U,...,2U,2U,...$, that is, $U_i=U_\textmd{left}=U$ and
$U_i=U_\textmd{right}=2U$.

\section{Conclusions}
\label{section.conclusions}
In conclusion, we have studied the quantum phases of interacting
phonons in ion traps. The superfluid--Mott insulator quantum phase
transition can be detected by the evolution of the phonon density profile, as
well as by the divergence of the correlation length near the quantum
critical point. Although boundary effects are important, specially in
the case of a Coulomb chain of ions, correlation functions show a
similar behavior as those of systems in the thermodynamical limit. For
example, Luttinger liquid theory gives an approximate description of
the algebraic decay of correlations in the superfluid regime.

We have also shown that the ability to control phonon--phonon
interactions allows us to study a variety of situations like
attractive interactions, where a phase with large phonon number
fluctuations takes place. The ability to tune locally the value of the
on--site interactions also leads to the realization of new exciting
models, where the degeneracy of the classical ground state can be
tuned by choosing properly the value of the phonon--phonon interactions.

Work supported by the Deutscher Akademischer Austausch Dienst, SCALA and
CONQUEST projects and the DFG.


\begin{thebibliography}{99}
\bibitem{note1} Note that long--range order is destroyed by quantum
  fluctuations even at zero temperature in one spatial dimension. However we
  use here the usual terminology in lattice models, and choose the term
  ``superfluid phase'' for the quantum phase $U < U_c$ of the
  Bose--Hubbard model. This phase does not show long--range order, but
  algebraic decay of the phase correlations.
\end{thebibliography}
\end{document}